\documentclass[prb, twocolumn, showpacs, superscriptaddress,longbibliography, floatfix]{revtex4-1}
\usepackage{amsmath, amsfonts, amssymb, mathrsfs}
\usepackage{graphicx, subfigure, color}
\usepackage{bm}
\usepackage{braket}
\usepackage{epstopdf}
\usepackage[normalem]{ulem} 
\usepackage{color}
\usepackage[breaklinks]{hyperref}
\usepackage{lipsum}
\usepackage[colorinlistoftodos]{todonotes}
\usepackage{xcolor}
\usepackage{enumerate}
\definecolor{grey}{rgb}{.6,.6,.6}

\hypersetup{
   colorlinks,
   citecolor=blue,
   filecolor=blue,
   linkcolor=blue,
   urlcolor=blue
}

\graphicspath{Graphes}

\begin{document}

\title{\texorpdfstring{$4\pi$ and $8\pi$ dual Josephson effects induced by symmetry defects}{}}

\author{Corneliu Malciu}
\affiliation{Laboratoire de physique de l'\'Ecole Normale Sup\'erieure, PSL Research University, CNRS, Universit\'e Pierre et Marie Curie-Sorbonne Universit\'es, Universit\'e Paris Diderot-Sorbonne Paris Cit\'e, 24 rue Lhomond, 75231 Paris Cedex 05, France}

\author{Leonardo Mazza}
\affiliation{LPTMS, UMR 8626, CNRS, Univ. Paris-Sud, Universit\'e Paris-Saclay, 91405 Orsay, France} 

\author{Christophe Mora}
\affiliation{Laboratoire de physique de l'\'Ecole Normale Sup\'erieure, PSL Research University, CNRS, Universit\'e Pierre et Marie Curie-Sorbonne Universit\'es, Universit\'e Paris Diderot-Sorbonne Paris Cit\'e, 24 rue Lhomond, 75231 Paris Cedex 05, France}

\begin{abstract}
In topological insulator edges,
the duality between the Zeeman field orientation and the proximitized superconducting phase has been recently exploited to predict a magneto-Josephson effect with a 4$\pi$ periodicity. 
We revisit this latter  Josephson effect in the light of this duality
and show that the same 4$\pi$ quantum anomaly occurs when bridging two spinless Thouless pumps to a p-wave superconducting region
that could be as small as a single and experimentally-relevant superconducting quantum dot - a point-like defect. 
This interpretation as a dual Josephson effect never requires the
presence of Majorana modes but rather builds on the topological properties of adiabatic quantum
pumps with $\mathbb Z$ topological invariants. It allows for the systematic construction of dual Josephson
effects of arbitrary periodicity, such as  $4 \pi$ and $8 \pi$, by using point-like defects whose symmetry
differs from that of the pump, dubbed symmetry defects. 
Although adiabatic quantum pumps are typically discussed via mappings to two-dimensional geometries, we show that
this phenomenology does not have any counterpart in conventional two-dimensional systems.
\end{abstract}

\date{\today}
\maketitle

\section{Introduction}

The understanding of topological phases of matter in terms of bulk symmetries allowed for a full classification of gapped non-interacting electronic systems. The ten Altland-Zirnbauer symmetry classes~\cite{Altland_Zirnbauer_1997}, identified by the presence or not of time reversal, particle-hole and chiral symmetry, define a periodic table of topological phases which provides the type of topological invariant -- namely $\mathbb{Z}$, $\mathbb{Z}_2$ or trivial -- expected for a given spatial dimension~$d$~\cite{Schnyder_Ryu_2008, Schnyder_Ryu_2009, Kitaev_2009}. Defects in such topological materials host protected subgap states, whose number -- $\mathbb{Z}$, $2 \mathbb{Z}$, $\mathbb{Z}_2$ or zero -- is fully characterized by the bulk Hamiltonian on a region surrounding the defect~\cite{Teo_Kane_2010}. 

In this framework, adiabatic pumps can be understood in terms of the quantized charges that they pump towards a defect, and they fall under the same classification~\cite{Teo_Kane_2010, Zhang_Kane_2014_PRB}. The simplest example is that of a one dimensional Thouless pump~\cite{Thouless_1983} where a quantized charge is moved through the bulk by performing a non-trivial closed trajectory in the parameter space~\cite{Teo_Kane_2010, Zhang_Kane_2014_PRB, Marra_Citro_2015}. The number of electrons pumped through the bulk is related to a Chern number, {\it i.e.}  a $\mathbb{Z}$ topological invariant. Such devices have been realized experimentally in quasi-crystals~\cite{Kraus_Lahini_2012}, in superconducting systems~\cite{Gasparinetti_Solinas_2012} and with ultracold atoms~\cite{Lohse_Schweizer_2015, Nakajima_Tomita_2016, Lu_Schemmer_2016}.

Following the previous classification, adiabatic pumps can also be characterized by a $\mathbb{Z}_2$ index, corresponding to the pumping of a $\mathbb{Z}_2$ charge, namely parity. Examples of such devices are the $\mathbb{Z}_2$ adiabatic spin pump~\cite{Fu_Kane_2006} and the $4\pi$ Josephson effect in p-wave superconductors~\cite{Kitaev_2001, Kwon_Sengupta_2004, Fu_Kane_2009, Zhang_Kane_2014_PRB}. This latter case corresponds to an anomalous periodicity of the low-energy sector in a superconducting-insulating-superconducting (S-I-S) junction as the superconducting phase difference is adiabatically advanced. It is deeply related to the presence of edge Majorana modes~\cite{Alicea_2012, Leijnse_Flensberg_2012, Beenakker_2013, Stanescu_Tewari_2013, Elliott_Franz_2015, DasSarma_2015, Sato_Fujimoto_2016, Aguado_2017}, and it should be understood as a parity switch of these edge states induced by the pump cycle~\cite{Teo_Kane_2010}. Such $4\pi$ periodicities have been observed in several experiments on devices predicted to host edge Majorana modes~\cite{Rokhinson_Liu_2012, Bocquillon_Deacon_2016, Wiedenmann_Bocquillon_2016, Deacon_Wiedenmann_2017}; however, it is worth mentioning that this observation is not an unambiguous evidence of their presence~\cite{Sau_Berg_2012, Chiu_DasSarma_2018, Vuik_Nijholt_2018}. 

A few years ago, it was pointed out that a $4\pi$ Josephson effect could also be obtained in dual setups, namely I-S-I junctions, by winding the orientation of a magnetic field in the insulating regions~\cite{Meng_Shivamoggi_2012, Jiang_Pekker_2013, Pientka_Jiang_2013, Kotetes_Schon_2013}, thereby producing a dissipationless spin current. This effect, dubbed \textit{magneto-Josephson effect}, has been put forward for several platforms involving the edge states of topological insulators or Rashba nanowires. It is essentially the result of a duality in quantum spin Hall insulators between a proximitized superconducting pairing and a Zeeman field. The $4\pi$ periodicity of the dual magneto-Josephson effect is related to a change in a local degree of freedom, such as a parity or spin flip~\cite{Meng_Shivamoggi_2012},  similarly to the usual $4\pi$ Josephson effect. Notwithstanding its simplicity,  the magneto-Josephson effect poses a number of interesting theoretical questions: i) to which topological phenomenon can we ascribe the  magneto-Josephson effect? ii) could it appear with higher periodicities, e.g. $8 \pi$? iii) how does it fit into the known classification for adiabatic pumps (see e.g.~[\onlinecite{Teo_Kane_2010}])? The goal of this article is to give a comprehensive answer to these three questions.

The Zeeman insulating region is a specific example of a Thouless pump. We thus show that the magneto-Josephson effect can be recovered in a simpler setup with spinless Thouless pumps for the insulating regions where the winding of the magnetic field orientation is replaced by adiabatic pump cycles. Here neither magnetic fields nor spin currents are present, and thus we give to this phenomenon the generic name of \textit{dual Josephson effect.}
Since the phenomenon survives (and actually is enhanced) when the superconducting inner region is shrunk to a point-like defect, we can safely decouple it from the phenomenology of boundary Majorana modes\cite{Meng_Shivamoggi_2012}.

We observe that both Thouless pumps are characterized by a $\mathbb{Z}$ Chern number representing the number of electrons pumped towards the edge during each pump cycle. In the presence of superconductivity in the central region, the pumped electrons can be absorbed by pairs, thus leaving zero or one electron in the Kitaev chain after one pump cycle. A second pump cycle restores the original ground state which corresponds to an anomalous $4\pi$ periodicity. The Thouless pumps effectively behave as \textit{parity pumps}: the $\mathbb{Z}$ invariant is broken down to $\mathbb{Z}_2$ by the superconducting boundary. The $\mathbb{Z}_2$ topological invariant is thus imposed by the boundary and not by the bulk as in the standard  $4\pi$ Josephson effect. Dual Josephson effects with higher periodicities (such as $8\pi$) can be similarly obtained by considering local defects with extended symmetries.

Alternatively, one can view quantum pumps in one-dimension as effective two-dimensional systems by identifying the phase of the pump as a wavevector in a second fictitious dimension of space. In this language, the fermions pumped towards the edge are viewed as chiral edge states in 2D and they are topologically protected by the Chern number. Despite this analogy, the physics is different when edge perturbations are added. The superconducting edge defects that we consider are no longer local in the 2D picture and they allow for processes breaking momentum conservation. This results in a specific classification of edge perturbations in quantum pumps compared to that of 2D systems.

The magneto-Josephson effect is thus found to be a pumping phenomenon dictated by an interplay between the bulk topology and the symmetries of a local defect (dubbed \textit{symmetry defect}), so that the bulk-boundary correspondence is modified by the local properties of the edges. As such, it does not fit into the classification proposed in Ref.~\onlinecite{Teo_Kane_2010} and its theoretical analysis adds a novel piece in the generic characterization of adiabatic quantum pumps.

This article is organized as follows: in Sec.~\ref{sect:4pi_Josephson_and_dual} we use an effective low-energy continuous description to link the standard magneto-Josephson effect to the dual Josephson effect appearing in a junction composed of two 1D Harper wires sandwiching a Kitaev chain.
In Sec.~\ref{sect:symmetry_defects} we introduce the concept of point-like symmetry defects and show that the features of the dual Josephson effects survive when the superconducting Kitaev chain is replaced by a point-like defect (e.g. a superconducting quantum dot). We present an interpretation of the phenomenon in terms of an interplay between the symmetries of the Thouless pump and of the defect, and show that this picture can be generalized in order to engineer higher periodicities such as dual $8\pi$ Josephson effects.
In Sec.~\ref{sect:dual_effect_lattice_model} we discuss a lattice model for a junction with a dual Josephson effect of periodicity $4 \pi$.
Our conclusions are presented in Sec.~\ref{sect:conclusion}.

\section{\texorpdfstring{Dual Josephson effect: a low-energy description}{}}\label{sect:4pi_Josephson_and_dual}

\subsection{Lattice models of a Thouless pump and of a Kitaev chain}\label{sect:models}

\begin{figure}
    \centering
    \includegraphics[width = .55\linewidth]{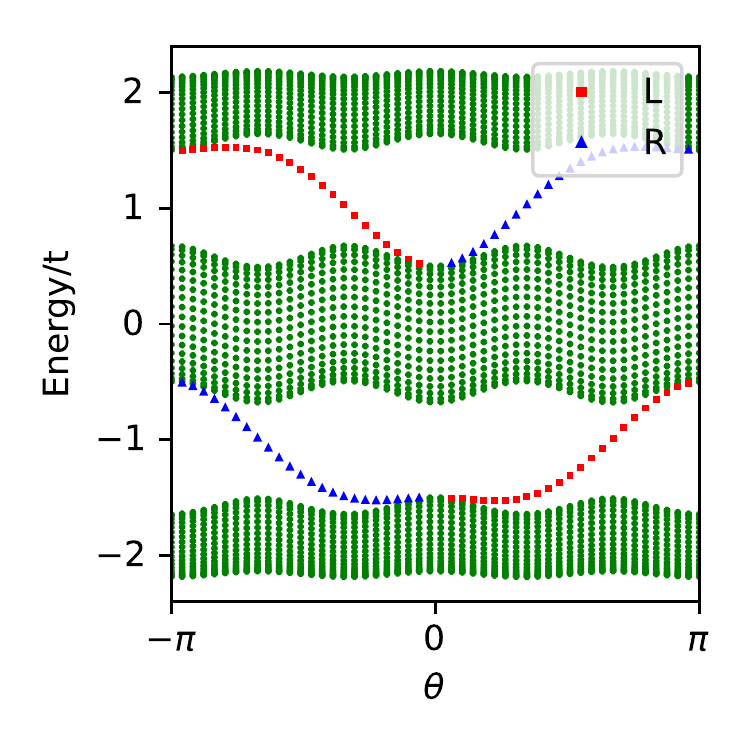}
    \includegraphics[width = .4\linewidth]{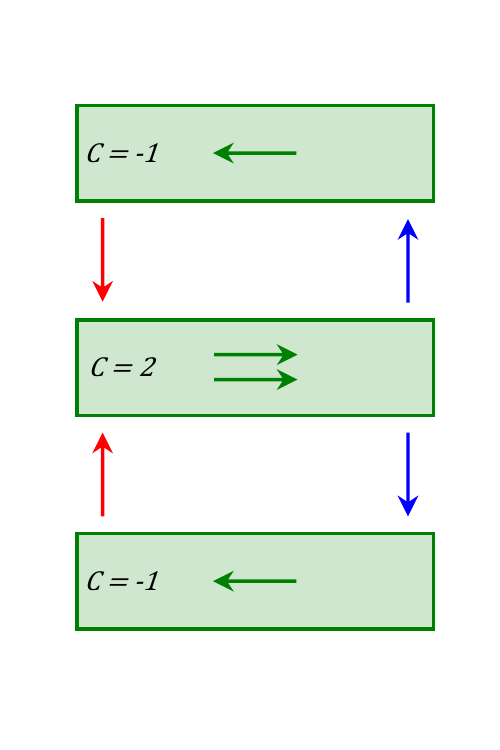}
    \caption{
    \textbf{Left panel:} Spectrum of the 1D Harper model~\eqref{eq:H_pump_micro} as $\theta$ is winded. Electrons in the upper and lower bands are pumped one site to the left; those in the central band are pumped two sites to the right. For open boundary conditions, electrons are transferred between bands at the left (L) and right (R) edges of the system. The plot is obtained for $V_0 = t$, $\mu = 0$ and a size $L = 59$. 
    \textbf{Right panel:} Scheme of the motion of the electrons during an adiabatic pumping cycle. The number of electrons pumped through the bulk corresponds to the Chern number ($C$) of each band.
    }
    \label{fig:Pump_spectrum}
\end{figure}

We first review several lattice models for a Thouless pump and introduce the Kitaev chain for the superconducting region. This gives us a microscopic formulation of our problem where the symmetries can be addressed in great detail. The Thouless pump that we will be the central part of this paper is a 1D Harper model~\cite{Harper_1955, Aubry_Andre_1980, Han_Thouless_1994, Marra_Citro_2015}. Its Hamiltonian reads:
\begin{align}\label{eq:H_pump_micro}
    H_{\text{Harp}} = \sum_j \bigg\{ &\left( -t \hat{a}^\dagger_j \hat{a}_{j+1} + \text{h.c.} \right) \\
    \nonumber
    &- \left(V_0 \cos\left(\frac{2\pi j}{3} + \theta\right) + \mu \right)\, \hat{a}^\dagger_j \hat{a}_{j} \bigg\},
\end{align}
where $\hat a_j^{(\dagger)}$ are fermionic annihilation (creation) operators.
It describes a tight-binding model with hopping amplitude $t$, chemical potential $\mu$, and an additional potential $V_0$ with periodicity of three sites modulated by a phase $\theta$; it is a three-band model. As $\theta$ is winded from $-\pi$ to $\pi$, electrons in the upper and lower bands are adiabatically moved one site to the left while those in the central band are moved two sites to the right. The electrons which reside at the edges cannot accumulate and they are transferred between bands by crossing the energy gaps as illustrated in Fig.~\ref{fig:Pump_spectrum}. 

The numbers of electrons crossing the band gaps during each pumping cycle are topologically protected. The topological origin is best revealed by considering $\theta$ as the momentum along an artificial dimension: the model then exactly maps onto a 2D Harper-Hofstadter model with flux $1/3$~\cite{Harper_1955, Hofstadter_1976}, as reviewed in Appendix~\ref{App:Thouless:Pumps}. The pumped electrons are then simply given by comparing the Chern numbers of each band in the fictitious 2D model. It is important to emphasize that the band crossings are robust against perturbations, such as disorder or changes of parameters, due to their topological character.

\begin{figure}
    \centering
    \includegraphics[width = \linewidth]{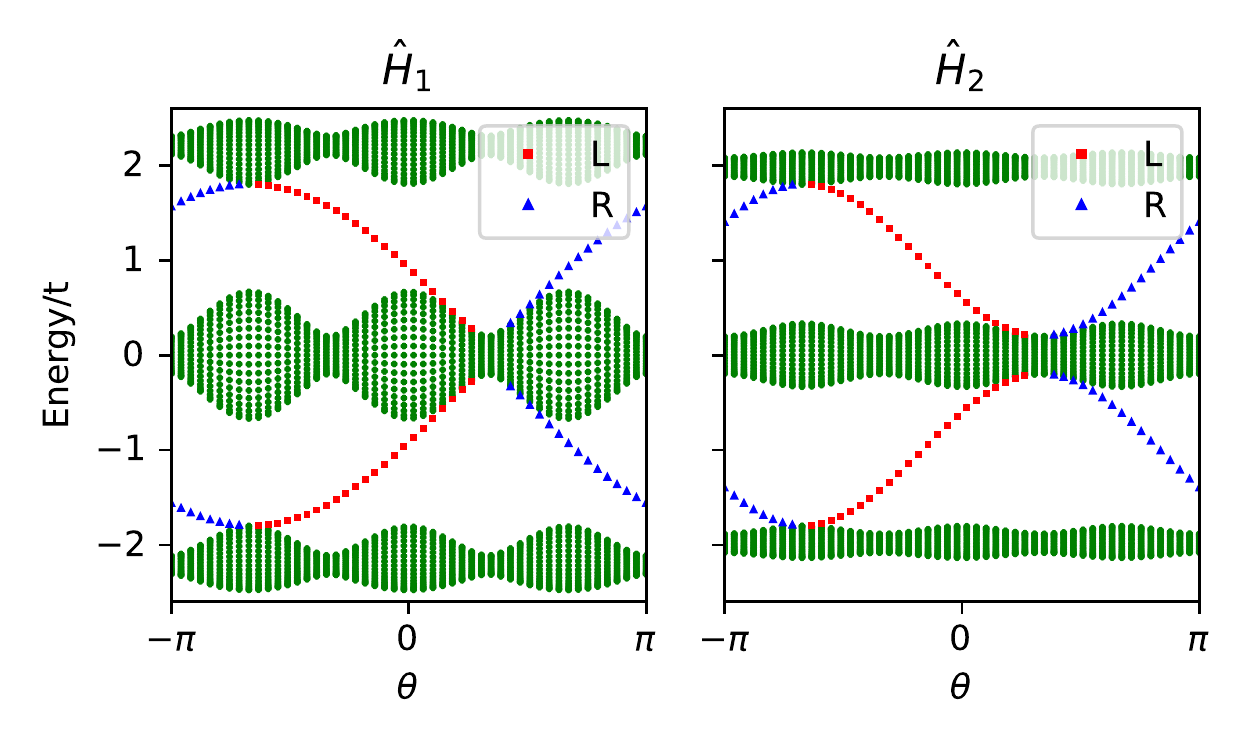}
    \caption{Spectrum of Hamiltonians~\eqref{eq:H_pump_micro_phase} and~\eqref{eq:H_pump_micro_ampl} as $\theta$ is winded. Both systems feature the same phenomenology as Hamiltonian~\eqref{eq:H_pump_micro} with states crossing the gap at the left (L) and rights (R) edges of the system (see Fig.~\ref{fig:Pump_spectrum}). The plot are obtained for $t' = 0.8t$, $\mu = 0$ and a size $L = 59$.}
    \label{fig:Pump_spectrum_2}
\end{figure}

Hamiltonian~\eqref{eq:H_pump_micro} has been implemented in quasi-crystals~\cite{Kraus_Lahini_2012} and ultracold atomic systems~\cite{Aidelsburger_Atala_2013, Miyake_Siviloglou_2013}; its experimental relevance motivates its choice.
Note however that the potential with periodicity of three sites could be replaced by i) a hopping phase with periodicity of three sites or ii) hopping amplitude with periodicity of three sites without changing the phenomenology. 
These models are described by
Hamiltonians $H_1$ and $H_2$ respectively (the chemical-potential term is omitted for brevity):
\begin{subequations}
\begin{align}
    \label{eq:H_pump_micro_phase}
    H_1 &=  \sum_{j} \left\{ - \left( t + t' e^{i\left(\frac{2 \pi j}{3} + \theta\right)}\right) \hat{a}^\dagger_j \hat{a}_{j+1} + \text{h.c.} \right\}\\
    \label{eq:H_pump_micro_ampl}
    H_2 &=  \sum_{j} \left\{ - \left( t + t' \cos\left(\frac{2 \pi j}{3} + \theta\right)\right) \hat{a}^\dagger_j \hat{a}_{j+1} + \text{h.c.} \right\},
\end{align}
\end{subequations}
where $t$ is a uniform hopping amplitude, $t'$ is the alternating hopping amplitude, and $\theta$ is a phase. 
The lowest band of these two Hamiltonians has a Chern number $C = -1$ (for Hamiltonian $H_2$, this is true if $t'<4t$), and they therefore act as Thouless pumps. Their mapping onto a two dimensional Hamiltonian is discussed in Appendix~\ref{App:Thouless:Pumps}, and their spectral flow with $\theta$ is shown in Fig.~\ref{fig:Pump_spectrum_2}.

The Kitaev chain~\cite{Kitaev_2001} is a 1D p-wave  superconductor whose topological phase exhibits edge Majorana modes. It can be written in terms of spinless fermions as:
\begin{equation}\label{eq:H_K_micro}
    H_K = \sum_j \left\{ \left( -t \hat{a}^\dagger_j \hat{a}_{j+1} + i \Delta_0 \hat{a}_j \hat{a}_{j+1} + \text{h.c.} \right) - \mu \, \hat{a}^\dagger_j \hat{a}_{j} \right\},
\end{equation}
where $\hat a_j^{(\dagger)}$ are fermionic annihilation (creation) operators, $t$ is a uniform hopping term, $\Delta_0 = |\Delta_0|e^{-i\phi}$ is the superconducting pairing term and $\mu$ the chemical potential. For $|\mu|<2t$ and $\Delta_0 \neq 0$, subgap Majorana modes with exponentially small energy appear at both ends of the chain. They are responsible for the $4\pi$ Josephson effect~\cite{Kitaev_2001, Fu_Kane_2006, Lutchyn_Sau_2010} and allow one to understand the Kitaev chain as a parity pump~\cite{Teo_Kane_2010}. 

\subsection{Low-energy duality and magneto-Josephson effect}\label{sect:duality}

An interesting duality between the 1D Harper model~\eqref{eq:H_pump_micro} and the Kitaev chain~\eqref{eq:H_K_micro} arises at low energy. In order to discuss this regime, we consider the continuum limit of the two models, at filling $1/3$ for the model~\eqref{eq:H_pump_micro} and half-filling for Eq.~\eqref{eq:H_K_micro}. By linearizing around the two Fermi points, we arrive at the two Hamiltonians written in terms of the chiral left and right movers (in the basis $\{\psi_R, \psi_L, \psi_L^\dagger, - \psi_R^\dagger\}$),
\begin{subequations}
\begin{align}
    \label{eq:H_pump_continuum}
    h_{\text{Harp}} &= -i v_F \partial_x \sigma_z\tau_z - V \left(\cos \theta \sigma_x - \sin \theta \sigma_y \right),\\[2mm]
    \label{eq:H_kitaev_continuum}
    h_K &= -i v_F \partial_x \sigma_z\tau_z  - \Delta \left(\cos \phi \tau_x - \sin \phi \tau_y \right);
\end{align}
\end{subequations}
where $v_F \sim t$, $V \sim V_0$ and $\Delta \sim \Delta_0$ at weak coupling. Here, $\sigma_i$ ($\tau_i$) are the Pauli matrices acting on the chiral (particle-hole) sectors. We note that the Hamiltonian~\eqref{eq:H_pump_continuum} is also the low-energy limit of the Rice-Mele model~\cite{Rice_Mele_1982,Teo_Kane_2010}. Remarkably, these two Hamiltonians are mapped onto each other by the duality transformation
\begin{equation}\label{eq:duality_mapping}
    \left(\Psi_L, \Psi_L^\dagger\right)  \leftrightarrow  \left(\Psi_L^\dagger, \Psi_L\right)
    \qquad 
    \left(V, \theta\right) \leftrightarrow \left(\Delta, \phi \right),
\end{equation}
which acts on a single chiral sector. This mapping is only valid at low energy and does not pertain  to the original models~\eqref{eq:H_pump_micro} and~\eqref{eq:H_K_micro}. In particular, the phases $\phi$ and $\theta$ have a different behavior: whereas $\phi$ can be removed from Kitaev's Hamiltonian by a simple $U(1)$ gauge transform,  the entire spectrum of the 1D Harper model depends on $\theta$. 

The two low-energy models~\eqref{eq:H_pump_continuum} and~\eqref{eq:H_kitaev_continuum} are precisely those of the insulating and superconducting regions discussed in Refs.~\onlinecite{Jiang_Pekker_2013, Pientka_Jiang_2013}
to describe the magneto-Josephson effect. We follow their presentation and introduce the one-dimensional edge state of a topological insulator with the chiral and spin-polarized fields $\psi_\downarrow (x), \psi_\uparrow (x)$. Adding superconductive pairing by proximity and applying a Zeeman field $B$, one finds the Hamiltonian
\begin{align}
    \label{eq:H_MJ}
    h_{\text{MJ}} = & -i v_F  \partial_x \sigma_z\tau_z + B \left(\cos \theta \sigma_x - \sin \theta \sigma_y \right) \\[1mm]
    \nonumber
    &+  \Delta \left(\cos \phi \tau_x - \sin \phi \tau_y \right),
\end{align}
written in the basis $\{\psi_\uparrow, \psi_\downarrow, \psi_\downarrow^\dagger, - \psi_\uparrow^\dagger\}$. This Hamiltonian reduces to Eqs.~\eqref{eq:H_pump_continuum} and~\eqref{eq:H_kitaev_continuum}  in the two limiting cases $\Delta=0$ and $B=0$ if we identify $B$ and $V$. $\phi$ is the phase of the superconductor, $\theta$ refers to the orientation of the magnetic field $B$. The model~\eqref{eq:H_MJ} exhibits two gapped phases: a topological superconducting phase (or $\Delta$ phase) when $\Delta$ is large, and a topological magnetic phase (or $B$ phase) when $B$ dominates. Majorana zero modes bind to domain walls separating these two phases. The configuration $\Delta-B-\Delta$, where two superconducting regions surround an insulating magnetic one, exhibits a $4\pi$ Josephson effect as the phase difference between the superconductors is winded. By contrast, the dual configuration $B-\Delta-B$ generates where a $4\pi$-periodic spin current when the relative orientation of the two Zeeman fields is rotated: this is the magneto-Josephson effect.

\begin{figure}
    \centering
    \includegraphics[width = \linewidth]{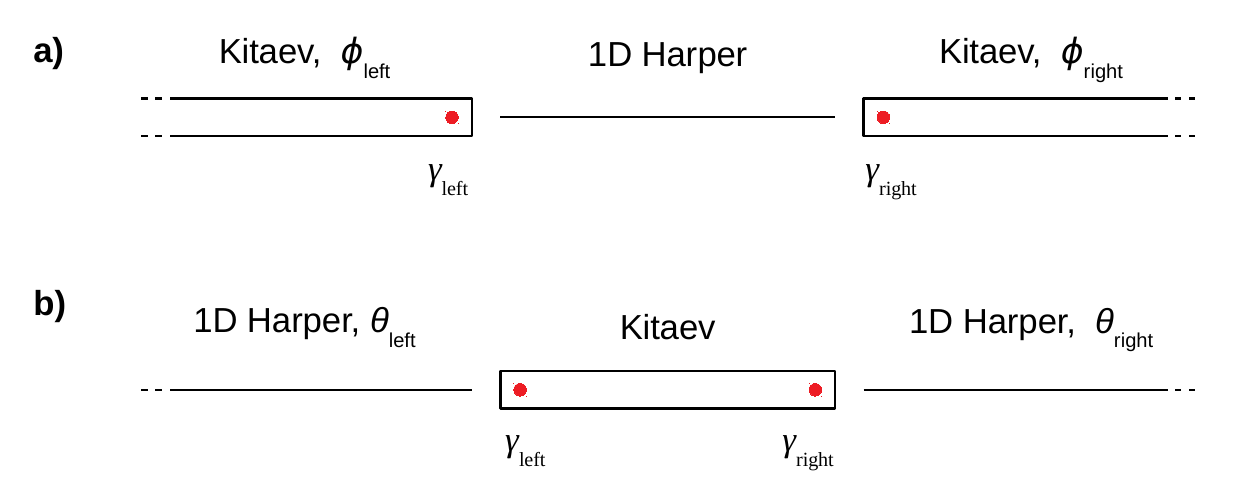}
    \caption{
    \textbf{a)} A finite 1D Harper model of length $L$ sandwiched between two semi-infinite Kitaev chains with superconducting phases $\phi_{\rm left, right}$. Majorana modes lie at both interfaces (at $x = 0,L$), and they hybridize through 1D Harper model with an exponentially small overlap. As $\phi_{\rm left} - \phi_{\rm right}$ is advanced by $2\pi$, the parity of the subgap state is changed; the system only recovers its initial state after a second cycle (see text).
    \textbf{b)} A finite Kitaev chain of length $L$ sandwiched between two semi-infinite 1D Harper models with phases $\theta_{\rm left, right}$. Majorana modes lie at both interfaces (at $x = 0,L$), and they hybridize through the Kitaev chain with an exponentially small overlap. As $\theta_{\rm left} - \theta_{\rm right}$ is advanced by $2\pi$, a single electron is injected into the superconducting region and switches its parity; the system only recovers its initial state after a second cycle (see text).
    }
    \label{fig:scheme_models}
\end{figure}

Keeping in mind the analogy with the Josephson effect in quantum spin Hall systems, we return to the Harper and Kitaev models, whose low-energy models are given in Eqs.~\eqref{eq:H_pump_continuum} and~\eqref{eq:H_kitaev_continuum}, and further discuss the two dual Josephson anomalies. For concreteness, we consider a setup of two semi-infinite Kitaev chains separated by an insulating Harper wire of length $L$, as sketched in Fig.~\ref{fig:scheme_models}~a), similar to $\Delta-B-\Delta$. The Majorana bound states $\hat\gamma_{\rm left}$ and  $\hat\gamma_{\rm right}$ located at the interfaces pair with an exponential overlap and form a subgap state with the Josephson Hamiltonian~\cite{Meng_Shivamoggi_2012,Jiang_Pekker_2013}
\begin{equation}\label{josephsonhamil}
 H_{Josephson} = i \hat\gamma_{\rm left} \hat\gamma_{\rm right} \Gamma \cos \left(\frac{\phi_{\rm left} - \phi_{\rm right}}{2}\right).
\end{equation}
The corresponding eigenenergies depend on the parity $i \hat\gamma_{\rm left} \hat\gamma_{\rm right} = \pm 1$ of the Majorana pair and  $\Gamma \simeq e^{-L}$  can be obtained within a simple transfer matrix approach as discussed in appendix~\ref{app:transfer_matrix}. The Hamiltonian~\eqref{josephsonhamil} describes a $4\pi$ Josephson effect~\cite{Kitaev_2001, Kwon_Sengupta_2004, Fu_Kane_2009, Zhang_Kane_2014_PRB}: as the phase difference $\phi_{\rm left} - \phi_{\rm right}$ is advanced by $2 \pi$, the fermion parity is flipped and the system recovers its initial state only after a second cycle.

The dual Josephson effect takes place when two 1D Harper models are separated by a Kitaev chain of length $L$. The corresponding geometry is represented in Fig.~\ref{fig:scheme_models}~b), similar to  $B-\Delta-B$. By duality between Eqs.~\eqref{eq:H_pump_continuum} and~\eqref{eq:H_kitaev_continuum}, the same low-energy Hamiltonian governs the subgap state,
\begin{equation}\label{dualjoseph}
 H_{dual} = i \hat\gamma_{\rm left} \hat\gamma_{\rm right} \Gamma \cos \left(\frac{\theta_{\rm left} - \theta_{\rm right}}{2}\right),
\end{equation}
which now involves the difference in the pumps phases, analogous to the magnetic field orientations. The same parity flip occurs with $\theta_{\rm left} - \theta_{\rm right}$ corresponding to a $4\pi$ quantum anomaly~\cite{Fu_Kane_2009} similar to the magneto-Josephson effect. These results can be recovered with a bosonization approach, as shown in Appendix~\ref{App:8PI:Bos}. Note however, that the spin current in the quantum spin Hall context, or $n_\uparrow - n_\downarrow$, is replaced in the Harper model by the difference  $n_R - n_L$ (number of left movers minus number of right movers) which has no lattice counterpart, and thus does not lead to a dissipationless current. 

This short overview of different known results adapted to our setup, with Kitaev and Harper regions, triggers already a few remarks. First, the magneto-Josephson effect can be seen as an adiabatic charge pumping phenomenon albeit characterized by a $\mathbb Z_2$ parity instead of the conventional $\mathbb Z$ Chern number of the Harper model. Then the size of the superconducting region can be reduced while keeping the description~\eqref{dualjoseph}, suggesting that the bulk Kitaev chain and their edge Majorana zero modes are not key elements for explaining the magneto-Josephson effect. These questions will be specifically addressed in the rest of this paper.

\section{Symmetry defects}\label{sect:symmetry_defects}

\subsection{\texorpdfstring{$\mathbb{Z}_2$ topology from a superconducting defect}{}}\label{sect:minimal_model}

The $4\pi$ dual Josephson effect predicted in Sec.~\ref{sect:duality} originates from a low-energy duality between the 1D Harper model and the Kitaev chain; in this setup, ground state oscillations are associated to parity flips of the Majorana edge modes appearing at the interfaces. In this section we show that this effect does not rely on the presence of Majorana edge modes: it can be generically engineered with a superconducting point-like defect hosting a low-energy subgap state. This is in stark contrast with the case of the standard Josephson effect of a S-I-S junction, where shrinking the  superconductors  to point-like regions completely spoils the phenomenon.

\begin{figure}
    \centering
    \includegraphics[width = .5\linewidth]{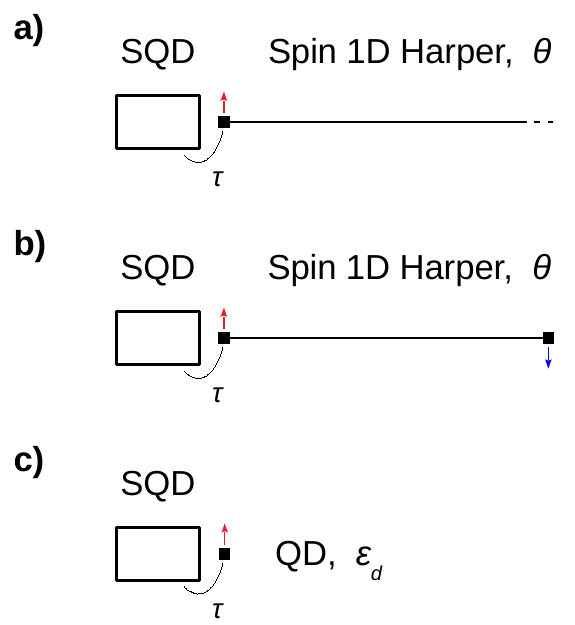}
    \caption{
    \textbf{a)} Superconducting quantum dot (SQD) connected to a semi-infinite spinful 1D Harper model~\eqref{eq:H_pump_micro_spin} by a hopping term $\tau$. As the phase $\theta$ of the adiabatic pump is winded, a boundary mode (black square) crosses the gap at the left edge of the system (red arrow). This setup is described by Hamiltonian~\eqref{eq:H_SQD_pump_micro_spin}.
    \textbf{b)} For a finite pump, another boundary mode crosses the gap at the right edge.
    \textbf{c)} Effective description of system \textit{a)}. The boundary mode crossing the gap can be replaced by a normal quantum dot (QD) whose energy is shifted. This corresponds to the effective Hamiltonian~\eqref{eq:H_SQD_pump_micro_spin_eff}.
    }
    \label{fig:scheme_eff_model_SQD}
\end{figure}

\begin{figure}
    \centering
    \includegraphics[width = .55\linewidth]{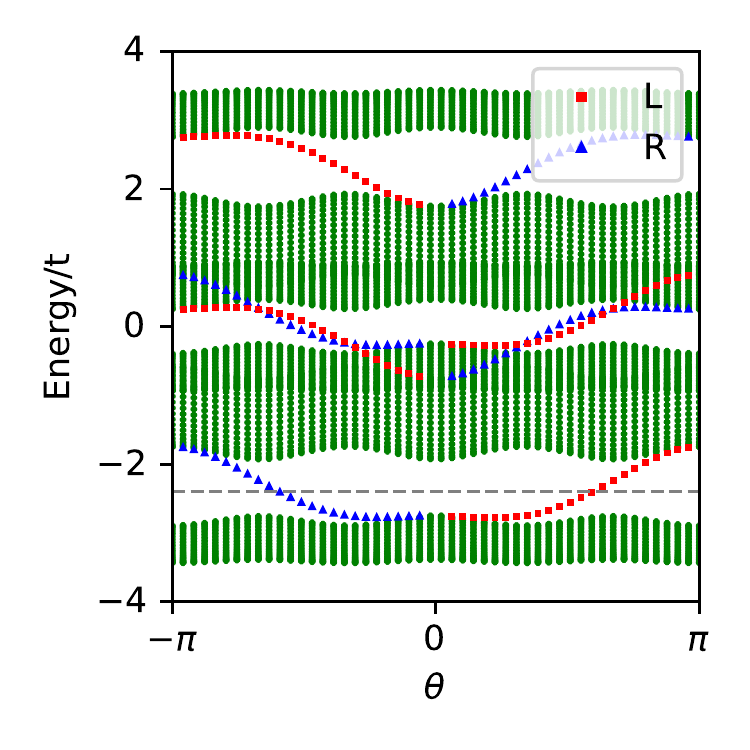}
    \raisebox{.27\height}{\includegraphics[width = .4\linewidth]{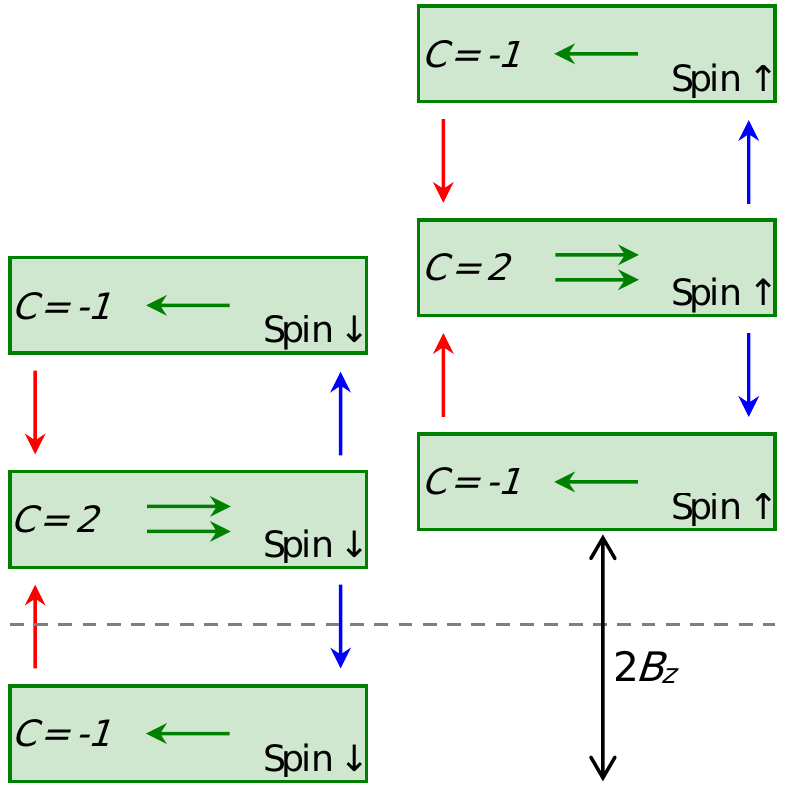}}
    \caption{
    \textbf{Left panel:} Spectrum of the spinful 1D Harper model~\eqref{eq:H_pump_micro_spin} as $\theta$ is winded. It corresponds to two replicas of Hamiltonian~\eqref{eq:H_pump_micro} that are split in energy by a magnetic field $B_z$. If the splitting is large enough, it is possible to set the chemical potential so that only the lowest band of the spin~$\downarrow$ copy is filled (grey dashed line); in this configuration, one electron is pumped towards the left edge during each cycle. The plot is obtained for $V_0 = t$,  $B_z = 1.25 t$, $\mu = 0$ and a size $L = 59$. 
    \textbf{Right panel:} Scheme of the motion of the electrons during an adiabatic pumping cycle. The number of electrons pumped through the bulk corresponds to the Chern number ($C$) of each band.
    }
    \label{fig:spectrum_sketch_BI3_spin}
\end{figure}

We illustrate this by connecting a Thouless pump to a superconducting quantum dot (SQD), which does not host Majorana zero modes. A sketch of the system is given in Figs~\ref{fig:scheme_eff_model_SQD}~a) and \ref{fig:scheme_eff_model_SQD}~b). For this example to be experimentally relevant, we consider a spinful system; the Hamiltonian of the Thouless pumps reads:
\begin{align}\label{eq:H_pump_micro_spin}
    H_{\text{Harp}}^{\text{S}} = &\sum_{j, \sigma = \uparrow\downarrow}\bigg\{ \left( -t \hat{a}^\dagger_{j, \sigma} \hat{a}_{j+1, \sigma} + \text{h.c.} \right) \\
    \nonumber
    &\qquad - \left(V_0 \cos\left(\frac{2\pi j}{3} + \theta\right) + \mu \right)\, \hat{a}^\dagger_{j, \sigma} \hat{a}_{j, \sigma} \bigg\}\\
    \nonumber
    &+ \sum_j \left\{B_z \hat{a}^\dagger_{j, \uparrow} \hat{a}_{j, \uparrow} - B_z \hat{a}^\dagger_{j, \downarrow} \hat{a}_{j, \downarrow}\right\}.
\end{align}
It corresponds to two spin-polarized copies of Hamiltonian~\eqref{eq:H_pump_micro} shifted in energy by a magnetic field $B_z$, that is with effective chemical potentials $\mu_\uparrow = \mu-B_z$ and $\mu_\downarrow = \mu+B_z$, as shown in Fig.~\ref{fig:spectrum_sketch_BI3_spin}. This model can in principle be implemented experimentally with an array of tunable quantum dots~\cite{Takakura, Baart, Fujita}. 

\begin{figure}
    \centering
    \includegraphics[width = \linewidth]{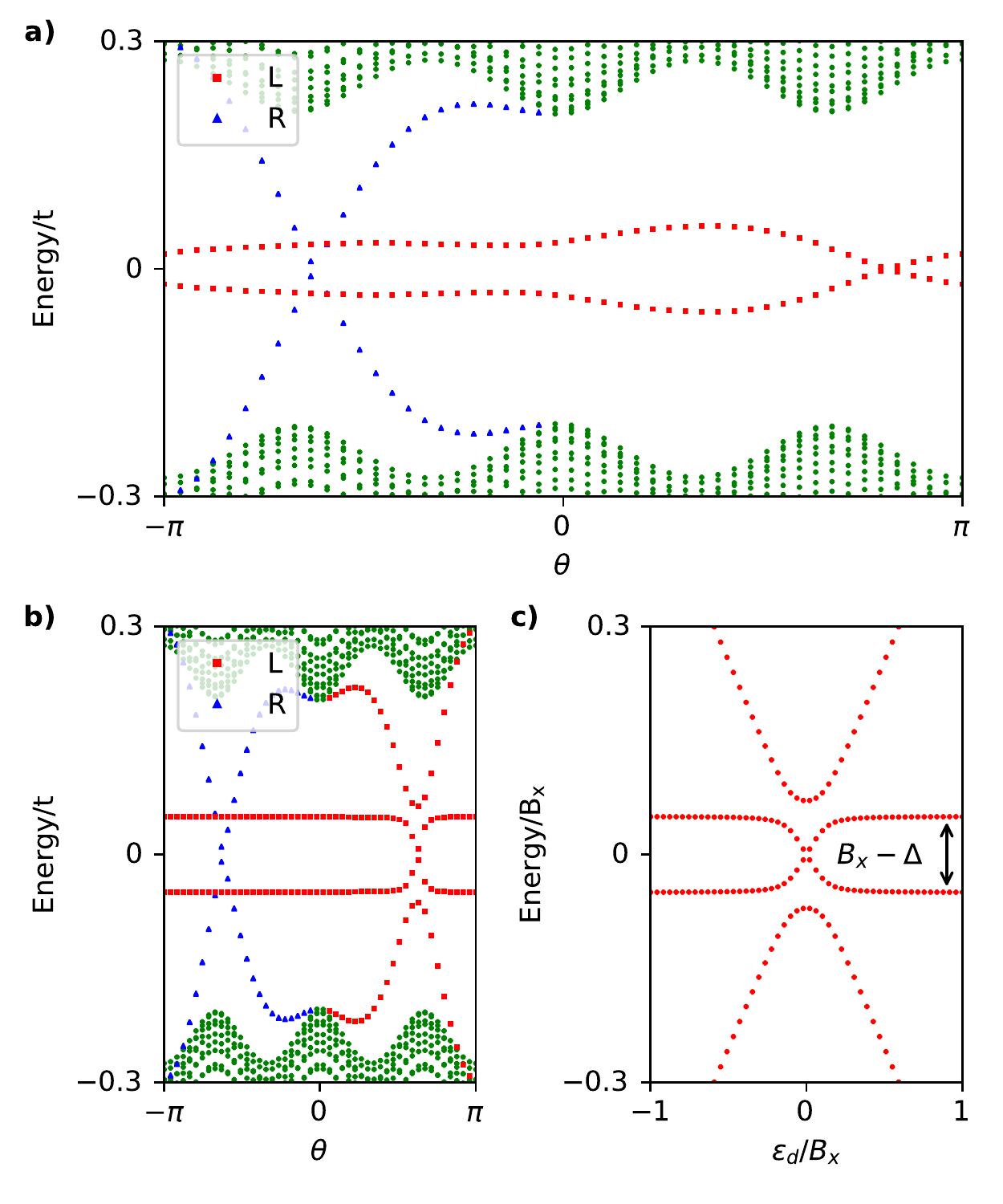}
    \caption
    {
    \textbf{a)} Energy spectrum of Hamiltonian~\eqref{eq:H_SQD_pump_micro_spin} as $\theta$ is winded from $-\pi$ to $\pi$. At the left edge (L), the dissipative band crossing in the absence of the SQD is turned by the SQD into a $4 \pi$ periodic spectral flow with a change of $\mathbb{Z}_2$ parity after each $2 \pi$ cycle. The  electron pumping, with a band crossing, at the right edge (R) can be discarded (by considering a semi-infinite limit or by adding a second SQD at the right edge as shown in Figs~\ref{fig:scheme_eff_model_SQD}~a) and \ref{fig:scheme_eff_model_SQD}~b). The plot was obtained for $V_0 = t$, $\mu = -2.3t$, $B_z = 1.2t$, $B_x = t$, $\Delta = 0.9t$, $\tau = 0.8t$ and $L = 59$. 
    \textbf{b)} Same as \textit{a)} for a weak coupling $\tau = 0.1t$.
    \textbf{c)} Effective model~\eqref{eq:H_SQD_pump_micro_spin_eff} of a linear fermionic mode (modelling the edge mode of the Thouless pump) connected to a SQD (see Fig~\ref{fig:scheme_eff_model_SQD}~c). This reproduces the low-energy physics plotted in \textit{b)} and shows how the SQD opens a gap in the edge mode dispersion relation.
    }
    \label{fig:spectrum_SQD_spin_BI3}
\end{figure}

If $B_z$ is strong enough, it is possible to set the chemical potential $\mu$ so that only the lowest band of the spin down copy is filled (see Fig.~\ref{fig:spectrum_sketch_BI3_spin}); that is, if $\theta$ is winded from $-\pi$ to $\pi$, Hamiltonian~\eqref{eq:H_pump_micro_spin} will only pump a spin down electron towards the left edge. We assume to be in this configuration, and we connect the left edge to a superconducting quantum dot described by the Hamiltonian:
\begin{equation}\label{eq:H_SQD}
    H_\text{SQD} = -\Delta \hat c_\uparrow \hat c_\downarrow + B_x \hat c^\dagger_\uparrow \hat c_\downarrow + \text{h.c.},
\end{equation}
where $\Delta$ is an s-wave superconducting pairing term, and where we have assumed for simplicity that the magnetic field orientation abruptly changes at the edge. This setup is sketched in Figs~\ref{fig:scheme_eff_model_SQD}~a) and \ref{fig:scheme_eff_model_SQD}~b).
A crucial feature of Eq.~\eqref{eq:H_SQD} is that it conserves the number of fermions modulo two so that the two sectors of even and odd parity are separated.
The entire setup is described by the Hamiltonian:
\begin{equation}\label{eq:H_SQD_pump_micro_spin}
    H = H_\text{SQD} + H_{\text{Harp}}^{\text{S}}  - \tau \left(\hat c^\dagger_\uparrow \hat a_{1, \uparrow} + \hat c^\dagger_\downarrow \hat a_{1, \downarrow} + \text{h.c.}  \right),
\end{equation}
where the hopping term $\tau$ between the SQD and the Thouless pump is spin-independent. The low-energy spectrum of the system, as $\theta$ is winded, is given in Fig.~\ref{fig:spectrum_SQD_spin_BI3}~a), and it exhibits a $4 \pi$
periodicity related to a change of $\mathbb{Z}_2$ parity. This shall be understood as electrons being pumped from the bulk into the SQD: during the first $2 \pi$ cycle, one spin down electron is injected into the SQD, flipping its parity and changing the low-energy state of the system; when a second cycle is performed, a second spin down electron is injected, but thanks to the spin-flipping term $B_x$ and the superconducting pairing $\Delta$, both electrons can escape into the superconducting reservoir, and the systems returns to its initial state. In the end, pumping electrons into the parity-conserving SQD amounts to pumping parity, and the $\mathbb{Z}$ invariant (Chern number) characterizing conventional electron pumping with dissipation is replaced by a non-dissipative adiabatic state evolution with a $\mathbb{Z}_2$ parity.

For small coupling $\tau \ll t$ between the SQD and the Thouless pump, the previous phenomenology is fully captured by discarding the bulk of the Thouless pump, and only considering an effective coupling to the left state (in red in Fig.~\ref{fig:spectrum_sketch_BI3_spin}) crossing the gap as $\theta$ is winded. This effective model is described by the Hamiltonian 
\begin{equation}\label{eq:H_SQD_pump_micro_spin_eff}
    H_{\text{eff}} = H_\text{SQD} + \varepsilon_d \hat d^\dagger_\downarrow \hat d_\downarrow - \tau \hat c^\dagger_\downarrow \hat d_{\downarrow} - \tau \hat d_{\downarrow}^\dagger \hat c,
\end{equation}
where $\hat d^{(\dagger)}$ are fermionic annihilation (creation) operators modelling the left edge state of the Thouless pump and $\varepsilon_d \sim \theta$ is a linear dispersion relation for this mode. This setup is sketched in Fig~\ref{fig:scheme_eff_model_SQD}~c) and the low-energy states of $H_{\text{eff}}$ as $\varepsilon_d$ crosses the gap are shown in Fig.~\ref{fig:spectrum_SQD_spin_BI3}~c) and reproduce very well the qualitative features of the full model shown in Fig.~\ref{fig:spectrum_SQD_spin_BI3}~b). In particular, one observes that the SQD opens a gap on the left edge state.

The gap opening is non-trivial since the left edge state is generally protected by a Chern number invariant and must cross the entire gap. The topological protection is broken here due to the presence of superconductivity, describing the fact the SQD absorbs the pumped electron. Nevertheless, what remains a topological invariant even with superconductivity is the $\mathbb{Z}_2$ parity of the number of times zero-energy is crossed as $\theta$ is winded by $2 \pi$ ($1$ in this particular case).

\subsection{Bulk-edge correspondence}\label{sect:bulk-boundary}

\begin{figure}
    \centering
    \includegraphics[width = \linewidth]{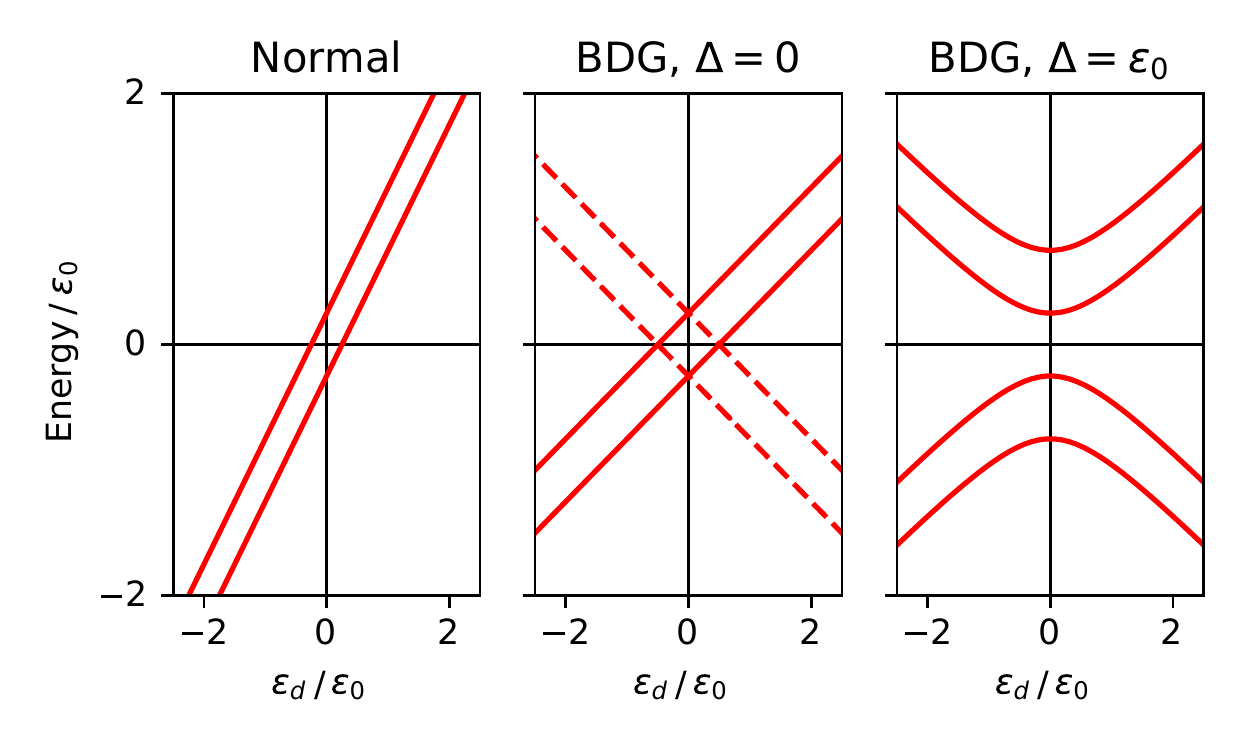}
    \caption{
    \textbf{Left panel:} Effective Hamiltonian~\eqref{eq:H_C2_eff} representing two edge modes crossing the gap of a spinful 1D Harper model; both zero-energy crossing are topologically protected, as energy levels shall be continuous and single-valued\cite{asboth_2016}.
    \textbf{Central panel:} Same Hamiltonian represented in the BDG formalism; each zero-energy crossing is protected by parity conservation.
    \textbf{Right panel:} A superconducting exchange term $\Delta$ can gap the pair of zero-energy crossings (see Eq.~\eqref{eq:H_C2_eff_S}).
    }
    \label{fig:C2_eff}
\end{figure}

In order to show that the parity of the number of zero-energy crossings is the invariant to be used in the presence of superconductivity, we now consider a case where there are two zero-energy crossings, and show that upon addiction of superconductivity, zero are left. As an example, let us consider Hamiltonian~\eqref{eq:H_pump_micro_spin} with $B_Z \approx 0$, so that both spin components are nearly degenerate (we take $B_Z \ne 0$ for better visibility). The lowest band of each spin component is filled, and two electrons (with spin $\uparrow, \downarrow$) cross the gap at the left edge as $\theta$ is winded. They can effectively be modelled by fermionic states $\hat d_{\uparrow, \downarrow}$ whose energies are tuned across the gap as:
\begin{equation}\label{eq:H_C2_eff}
    H_{\text{eff}} = \left(\varepsilon+\frac{\varepsilon_0}{2}\right) \hat d^\dagger_\uparrow \hat d_\uparrow + \left(\varepsilon-\frac{\varepsilon_0}{2}\right) \hat d^\dagger_\downarrow \hat d_\downarrow,
\end{equation}
where $\varepsilon_d$ crosses the gap when $\theta$ is winded, and $\varepsilon_0$ is the detuning between the two spin components.
We now add a superconducting pairing term $\Delta$ on the very left site of the Thouless pump, that is:
\begin{equation}\label{eq:S_coupling}
    H_S = - \Delta \hat a_{1,\uparrow} \hat a_{1,\downarrow} - \Delta \hat a^\dagger_{1,\downarrow} \hat a^\dagger_{1,\uparrow};
\end{equation}
the previous effective Hamiltonian becomes:
\begin{equation}\label{eq:H_C2_eff_S}
    H_{\text{eff}}^S =  \left(\varepsilon+\frac{\varepsilon_0}{2}\right) \hat d^\dagger_\uparrow \hat d_\uparrow + \left(\varepsilon-\frac{\varepsilon_0}{2}\right) \hat d^\dagger_\downarrow \hat d_\downarrow - \Delta \hat d_\uparrow \hat d_\downarrow - \Delta \hat d^\dagger_\downarrow \hat d^\dagger_\uparrow
\end{equation}
If $\Delta > \varepsilon_0/2$, the initial pair of zero-energy crossings is gapped, as shown in Fig.~\ref{fig:C2_eff}.

Summarizing, the number of topologically-protected zero-energy crossings changes from $\mathbb{Z}$ (the Chern number) to $\mathbb{Z}_2$ (the parity $\nu$ of the Chern number) in the presence of a superconducting defect. That is, the bulk-boundary correspondence is modified by the local symmetry properties of the edge, which hosts what we dub a \textit{symmetry defect}. Relying only on the Chern number of the filled bands does not provide a correct picture of the system. This $\mathbb{Z}_2$ index is already visible in Sec.~\ref{sect:minimal_model}: after an adiabatic cycle, the ground state parity is flipped (if $\nu = 1$) or not (if $\nu = 0$). As such, the low-energy degree of freedom of the SQD acts as a probe for counting the parity of the number of electrons injected by the Thouless pump.

\subsection{Synthetic dimensions}

\begin{figure}
    \centering
    \includegraphics[width = .65\linewidth]{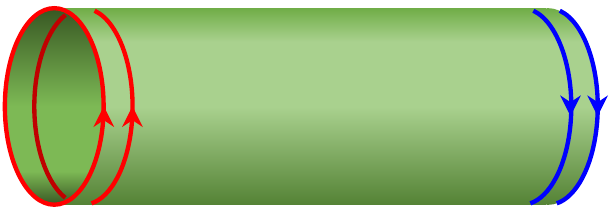}
    \caption{Mapping of the Thouless pump~\eqref{eq:H_pump_micro_spin} onto a 2D model by considering $\theta$ as the momentum along a periodic direction. The edge modes of the Thouless pump become chiral edge modes along this periodic direction; their dispersion relation is given by the spectrum of the pump at fixed $\theta$.}
    \label{fig:cylinder}
\end{figure}

The discussion of Thouless pumps generically relies on the interpretation of the parameter $\theta$ as the momentum along an additional synthetic dimension (see Sec.~\ref{sect:4pi_Josephson_and_dual} and Appendix~\ref{App:Thouless:Pumps} for details on the mapping). The spectrum of the 2D model at fixed momentum $\theta$ corresponds exactly to that of the Thouless pump for a fixed phase $\theta$, and the edge modes of the adiabatic pump become chiral edge modes propagating along the periodic dimension in the 2D system (see Fig.~\ref{fig:cylinder}).
It is thus worth investigating whether
the breakdown of $\mathbb{Z}$ to $\mathbb{Z}_2$ could also be observed in the number of chiral edges of a 2D system. 

In this context, the superconducting edge term introduced in Eq.~\eqref{eq:S_coupling} becomes:
\begin{equation}\label{eq:S_coupling_2D}
    H_S = \sum_\theta - \Delta \hat a_{1, \theta,\uparrow} \hat a_{1, \theta,\downarrow} - \Delta \hat a^\dagger_{1, \theta,\downarrow} \hat a^\dagger_{1, \theta,\uparrow},
\end{equation}
where $\hat a^{(\dagger)}_{1, \theta, \sigma}$ annihilates (creates) an electron with spin $\sigma$ and momentum $\theta$ along the periodic direction at the very left edge (site $x=1$). This corresponds to an exchange of Cooper pairs with non-zero total momentum, which is unrealistic in experimental implementations. The topological properties of the chiral edge modes can only be broken by adding non-local terms to the Hamiltonian. Such terms are possible in an adiabatic pump because there are no locality constraints along the synthetic dimension: terms that would be non-local along it can map onto local terms in real space, as shown by Eq.~\ref{eq:S_coupling} and Eq.~\ref{eq:S_coupling_2D}. We emphasize that the non-locality of edge interaction in the fictitious 2D system is a generic feature of such 1D-2D mapping\cite{Celi_Massignan_2014, Barbarino_Taddia_2015}.

\subsection{\texorpdfstring{$8\pi$ dual Josephson effect}{8 pi dual Josephson effect}}

\begin{figure}
    \centering
    \includegraphics[width = .5\linewidth]{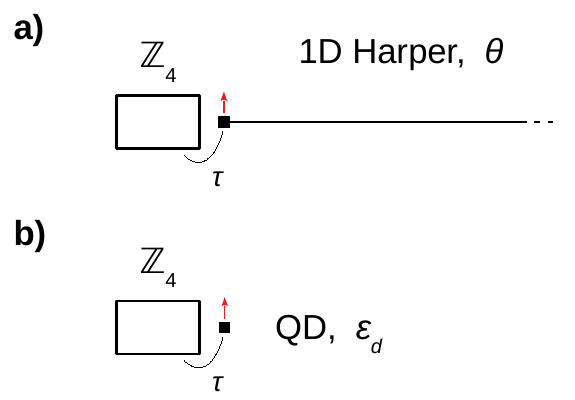}
    \caption{
    \textbf{a)} $\mathbb{Z}_4$ impurity~\eqref{eq:H_quadruplet} connected to a semi-infinite 1D Harper model by a hopping term $\tau$. As the phase $\theta$ of the adiabatic pump is winded, a boundary mode (black square) crosses the gap at the left edge of the system (red arrow).
    \textbf{b)} The boundary mode crossing the gap can effectively be described by a normal quantum dot (QD) whose energy is shifted. This corresponds to the effective Hamiltonian~\eqref{eq:H_Z4}.
    }
    \label{fig:scheme_eff_model_Z4}
\end{figure}

The idea that edge properties partly dictate the topological invariant can be further tested by considering two semi-infinite 1D Harper models separated by a spinless pairing model with $\mathbb{Z}_4$ symmetry, 
\begin{equation}\label{eq:H_quadruplet}
    H_{\Delta_4} = \sum_j  \left( -t \hat{c}^\dagger_j \hat{c}_{j+1} - \Delta_4  \hat{c}_j \hat{c}_{j+1} \hat{c}_{j+2}  \hat{c}_{j+3} + \text{h.c.} \right),
\end{equation}
namely the tight-binding model coherently exchanging quadruplets with a reservoir introduced in Ref.~\onlinecite{Mazza_Mora_2018}.
It is argued that it exhibits an $8 \pi$ periodic spectral flow as the phase difference of the pumps is winded, an effect that is dual to the $8 \pi$ Josephson effect identified in topological insulators and spin-orbit quantum wires~\cite{Zhang_Kane_2014_PRL,Orth_Tiwari_2015,Peng_VinklerAviv_2016,Chew_Mross_2018,calzona2018}. 
We revisit and derive this  $8 \pi$ dual Josephson effect with a rigorous bosonization analysis in appendix~\ref{App:8PI:Bos}. The $8 \pi$ periodicity is the result of energy crossings protected by the  $\mathbb{Z}_4$ parity symmetry of Hamiltonian~\eqref{eq:H_quadruplet} related to the conservation of the number of particles modulo $4$.

Following the spirit of Sec.~\ref{sect:minimal_model}, we can check that the same $8 \pi$ periodicity occurs when a single Harper model is connected to the model~\eqref{eq:H_quadruplet} having only a few sites, say $L=5$, as illustrated in Fig.~\ref{fig:scheme_eff_model_Z4}~a). The generalized parity $\hat Q = i^{\hat N}$ commutes with the Hamiltonian~\eqref{eq:H_quadruplet} and we are interested in the ground state in each sector of $\hat Q = \{\pm i, \pm 1\}$. The coupling to the Harper model mixes these four ground states. We expect that each adiabatic cycle where $\theta$ is advanced by $2 \pi$ injects one electron into the quadruplet model~\eqref{eq:H_quadruplet} and circulates the four ground states, leading to an $8 \pi$ periodic oscillation between the four low-lying states of the full model.

\begin{figure}
    \centering
    \includegraphics[width = \linewidth]{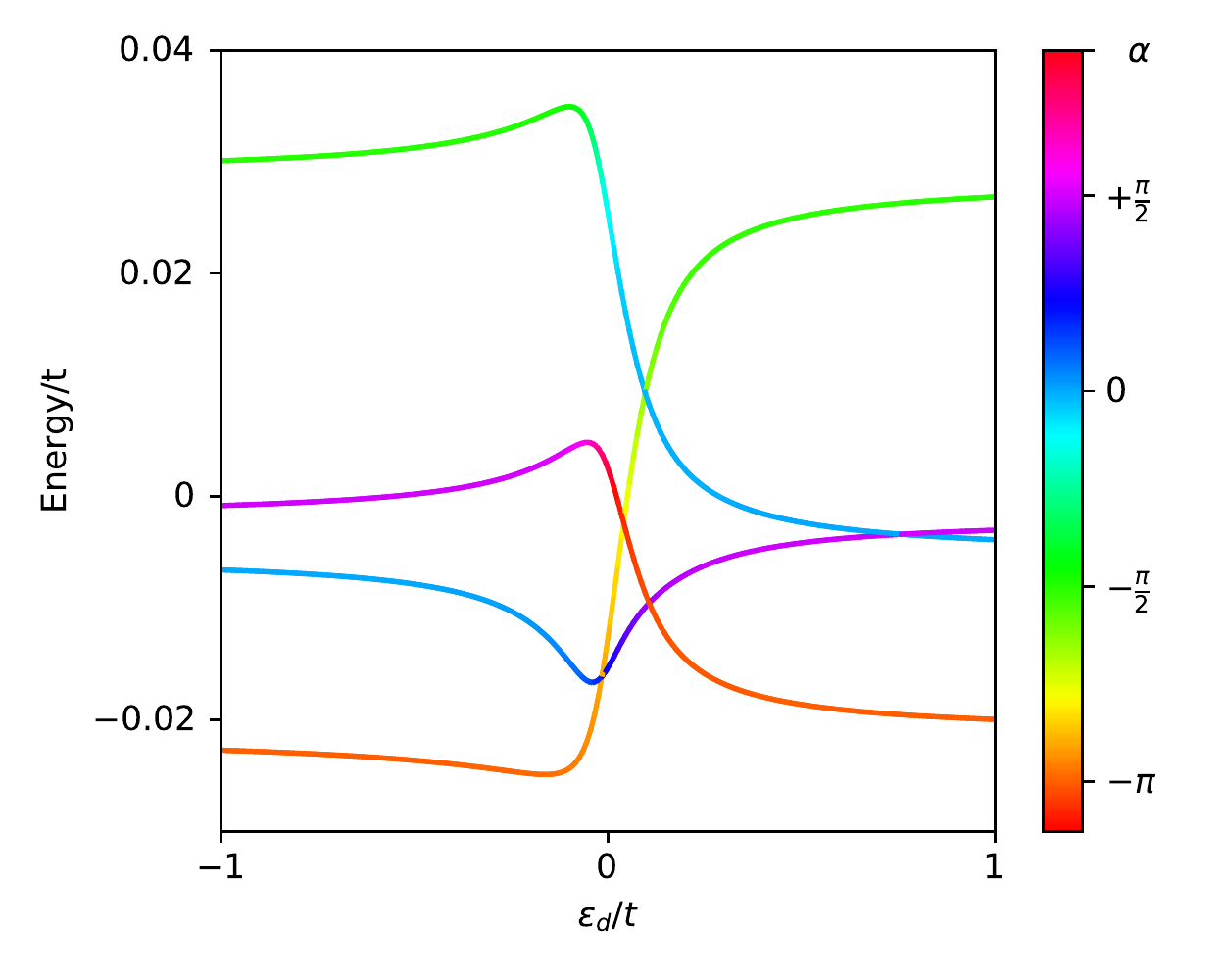}
    \caption{Many-body spectrum of Hamiltonian~\eqref{eq:H_Z4} as the edge state crosses the gap. The four low-energy states of the defect are interchanged with a $\mathbb{Z}_4$ periodicity, meaning that the system only recovers its initial state after four edge states cross the gap. Each of the low-energy states acquires a unit of generalized parity $\hat Q = i^{\hat N}$ during the injection of the electron (the color of the curves correspond to the phase $\alpha$ of $\hat Q$). The plot was obtained for $L=5$, $\mu = 0.05t$, $\Delta_4 = 1.75 t$ and $\tau = 0.1t$. For better visibility, the mean value of the low-energy subspace was subtracted.}
    \label{fig:spectrum_Quadruplet_Z4}
\end{figure}

To further simplify the analysis, we use an effective Hamiltonian similar to \eqref{eq:H_SQD_pump_micro_spin_eff} where the description of the Harper model is reduced to the boundary state crossing the gap during a cycle of $\theta$ (see Fig.~\ref{fig:Pump_spectrum}), which is modeled by a quantum dot whose energy $\varepsilon_d$ is tuned across zero energy. The effective low-energy Hamiltonian is
\begin{equation}\label{eq:H_Z4}
    H = H_{\Delta_4} - \tau \hat c^\dagger_L \hat{d} - \tau \hat d^\dagger \hat c_L + \varepsilon_d \hat d^\dagger \hat d,
\end{equation}
and the corresponding setup is sketched in Fig.~\ref{fig:scheme_eff_model_Z4}~b).
We diagonalize it with $\varepsilon_d$ tuned from $-10\tau$ to $+10\tau$  for a defect of size $L = 5$. The many-body spectrum of the defect has a four-fold low-energy subspace, which is rearranged with a $\mathbb{Z}_4$ periodicity as the quantum dot crosses the gap. The corresponding spectral flow of the four low-lying states is shown in Fig.~\ref{fig:spectrum_Quadruplet_Z4} where it is visible that the states have been interchanged. Each state acquires a unit of generalized parity so that this process of sweeping the quantum dot energy must be repeated four times to return to the initial state.

In terms of the original model with the Harper Thouless pump, each winding of $\theta$ by $2 \pi$ is equivalent to sweeping once the quantum dot energy across the gap so that the four-periodicity results in the announced $8 \pi$ dual Josephson effect. A simple picture also emerges if we focus on the edge states, which can be viewed as chiral propagating edge states in the synthetic dimension analogy of Sec.~\ref{sect:bulk-boundary}. The four-fermion pairing is able to gap those edge states four by four and thus breaks the $\mathbb{Z}$ Chern number down to a $\mathbb{Z}_4$ invariant which is the number of zero-energy crossing after one cycle (or the Chern number modulo $4$). The $8 \pi$ periodicity is then linked to this $\mathbb{Z}_4$ invariant.

\section{\texorpdfstring{$4 \pi$ dual Josephson effect: a lattice model}{4 pi dual Josephson effect: a lattice model}}\label{sect:dual_effect_lattice_model}

\begin{figure}
    \centering
    \includegraphics[width = \linewidth]{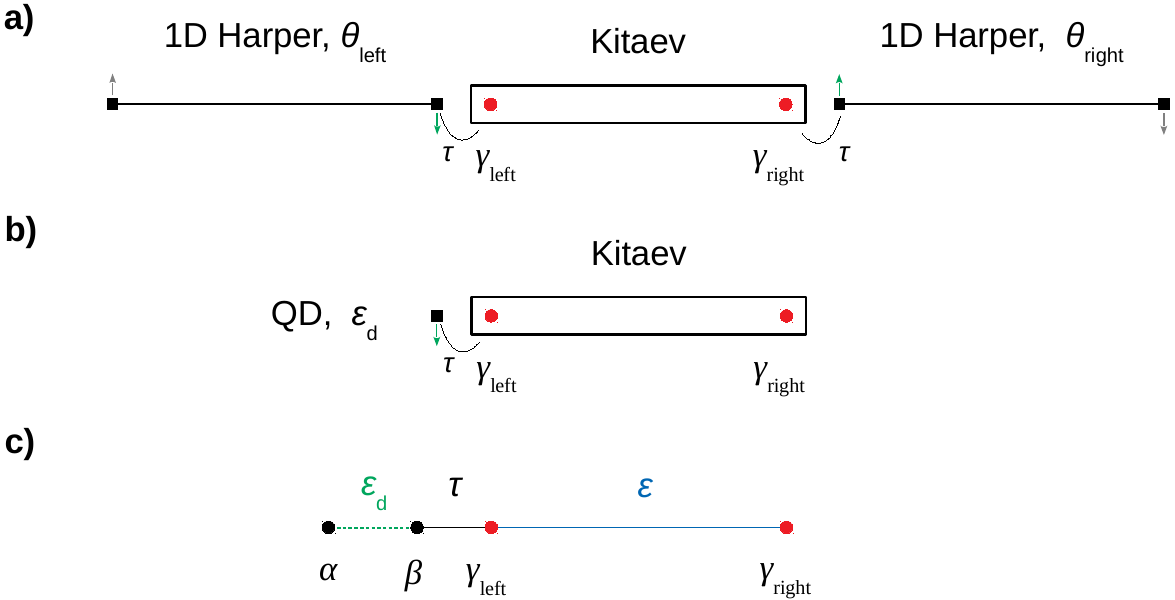}
    \caption{
    \textbf{a)} $H-K-H$ setup: two 1D Harper models are connected to a Kitaev chain of size $L$. As the phases $\theta_{\rm{left, right}}$ are winded, boundary modes (black square) crosses the gap at the edges of the adiabatic pumps.
    \textbf{b)} We focus only on the left Harper region. An effective model can be obtained by discarding the bulk of the adiabatic pump and replacing its boundary mode by a normal quantum dot (QD) whose energy $\varepsilon_d$ is shifted.
    \textbf{c)} By describing the QD in terms of Majorana modes $\hat\alpha, \hat\beta$, the effective model acquires the simple form given in Eq.~\eqref{eq:H_QD-K_eff}.
    }
    \label{fig:scheme_eff_model_mzm}
\end{figure}

Our derivation of the $4 \pi$ dual Josephson effect has been limited to two cases: (i) the continuum limit of two 1D Harper models surrounding a Kitaev chain, and (ii) a lattice Hamiltonian for a 1D Harper model connected to a SQD. For (i), a mapping to the physics of the magneto-Josephson effect in quantum spin Hall system has been established. For (ii), the physics can be reduced to the minimal model of Hamiltonian~\eqref{eq:H_SQD_pump_micro_spin_eff}, which exhibits the desired parity crossing. 
We now wish to explore a full lattice model, called $H-K-H$, with two 1D Harper models~\eqref{eq:H_pump_micro} connected via a hopping of amplitude $\tau$ to a Kitaev chain~\eqref{eq:H_K_micro} of size $L$ (we take the same hopping $t$ within the three subparts for simplicity); the setup is sketched in Fig.~\ref{fig:scheme_eff_model_mzm}~a).

\begin{figure}
    \centering
    \includegraphics[width = \linewidth]{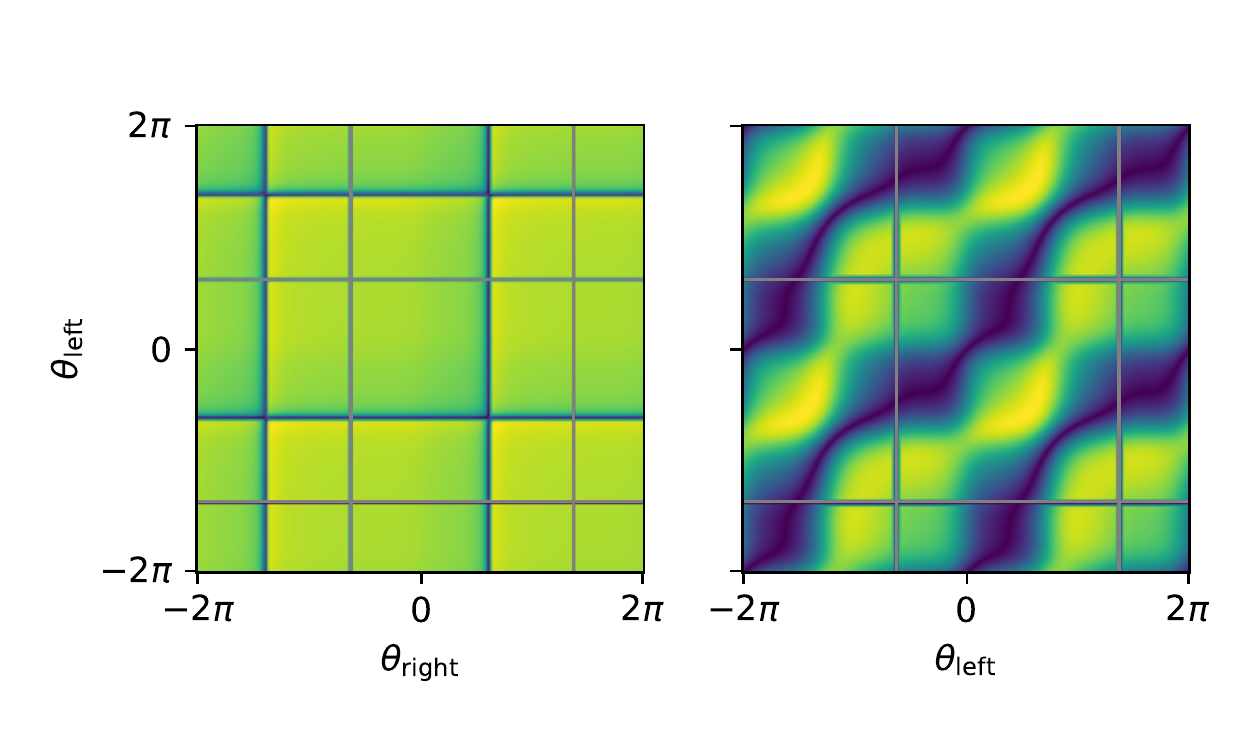}
    \caption{Low-energy spectrum of two 1D Harper models of length $L' = 59$ connected by a Kitaev chain of size $L = 12$. Blue regions correspond to zero energy, and therefore to a parity switch of the ground state. Grey lines correspond to zero energy levels originating from the states at the outer edges of the Thouless pumps and would not be present in an infinite system (see Fig.~\ref{fig:scheme_eff_model_mzm}~a); they shall be discarded for the discussion. \textbf{Left panel}: weak coupling regime ($\tau = 0.1t$) described by the effective Hamiltonian~\eqref{eq:H_QD-K_eff}; parity switches occur at fixed values of $\theta_{\rm left, right}$. \textbf{Right panel}: when approaching the conditions of the continuous model (here, $\tau = t$), parity switches are induced by the total charge at the boundaries of the Kitaev chain, and can no longer be predicted by a single phase $\theta_{\rm left, right}$; for $V_0, \Delta \ll t$ and $L, L'\gg 1$ they occur for $\theta_{\rm left} - \theta_{\rm right} = \pi \bmod 2\pi$ (see Eq.~\eqref{dualjoseph}). Both plots were obtained for $\Delta = 0.1t$ and $\mu = -1.09t$.}
    \label{fig:diag}
\end{figure}

We numerically solve this lattice model for several lengths $L$ and ratios $\tau/t$, and always recover the $4 \pi$ periodicity as function of either $\theta_{\rm left}$ or $\theta_{\rm right}$ (two examples are shown in Fig.~\ref{fig:diag}). 
Moreover, we find that the two cases (i) and (ii) are recovered as limiting cases. The continuum (i) is obtained by setting $t = \tau$ (perfect contacts), $V_0, \Delta_0 \ll t$ and a sufficiently large system. In this case, charges do not really accumulate on the left and right interfaces and the spectral flow is essentially a function of the difference $\theta_{\rm left} - \theta_{\rm right}$ in agreement with the continuum model (see Fig.~\ref{fig:diag}). 
The opposite limit (ii) is reached when $L=2$ for the Kitaev region and an exact mapping to the SQD holds by identifying the two sites operators $\hat a_{1,2}$ respectively with the spin projections  $\hat c_{\uparrow,\downarrow}$, and setting $t=B_x$.

Interestingly, for weak coupling between the regions $\tau\ll t$, the subgap physics can be captured by an effective description similar to Hamiltonian~\eqref{eq:H_SQD_pump_micro_spin_eff}. In this regime, charges accumulate at the two interfaces and the two spectral flows as a function of $\theta_{\rm left}$ and of $\theta_{\rm right}$ become independent (each of them exhibits the $4 \pi$ periodicity). This is illustrated in particular in Fig.~\ref{fig:diag} where the zero-energy crossings form perpendicular lines in the ($\theta_{\rm left}$, $\theta_{\rm right}$) plane. Thanks to this property, it is legitimate to discard one Harper region, as done in the case (ii), and study the spectral flow only as function of one phase  $\theta_{\rm left}$. Then, a low-energy description of the pumping can be built by focusing only on the edge state crossing the gap, that is modelled as a single level as in Sec.~\ref{sect:minimal_model}; this is illustrated in Fig.~\ref{fig:scheme_eff_model_mzm}~b).

We introduce the two Majorana bound states $\hat \gamma_{\rm left}$, $\hat \gamma_{\rm right}$ that are at the boundaries of the Kitaev chain and the two Majorana modes $\hat{\alpha}$ and $\hat{\beta}$ that are necessary to describe the edge mode of the Harper region (see Fig.~\ref{fig:scheme_eff_model_mzm}~c)).
The corresponding Hamiltonian reads:
\begin{equation}\label{eq:H_QD-K_eff}
    H_{\text{eff}} = i \left( \varepsilon_d \,\hat{\alpha} \hat{\beta} + \varepsilon \, \hat{\gamma}_{\rm left} \hat{\gamma}_{\rm right} + \tau \, \hat{\beta} \hat{\gamma}_{\rm left} \right),
\end{equation}
where $\varepsilon_d \sim \theta_{\rm left}$ accounts for the edge mode crossing the gap and $\varepsilon$ is the exponential energy splitting of the Kitaev chain. Note that, for simplicity, the QD is only coupled to the left Majorana mode of the Kitaev chain.

Hamiltonian~\eqref{eq:H_QD-K_eff} is also a low-energy description of a quantum dot connected to a Kitaev chain~\cite{Clarke_2017, Prada_San-Jose_2017, MMM_2018, Zeng_Moore_2018}, implemented experimentally~\cite{Deng_2016, Deng_2017}. It exhibits a zero-energy crossing at $\varepsilon_d = 0$ and an avoided crossing of the quantum dot energy level. As discussed in Ref.~[\onlinecite{MMM_2018}], it corresponds to the injection of an electron towards the Kitaev chain as the edge mode crosses zero energy, thereby flipping the fermion parity of the pair of Majorana edge modes. Again a second cycle is necessary to recover the initial state which elucidates the $4 \pi$ periodicity. This leads us to interpret the recent experimental studies in the light of a dual Josephson effect.

Summarizing,
the lattice $H-K-H$ model connects adiabatically the cases (i) and (ii), and thus demonstrates the common physical origin of their $4 \pi$ dual Josephson effects; this also rules out an explanation in terms of boundary Majorana modes.

\section{Conclusion}\label{sect:conclusion}

In this article, we revisited the magneto-Josephson effect and showed its appearance in a simpler model of junction where two semi-infinite Thouless pumps sandwich a superconducting nanowire; in this context, where magnetic fields are absent, we use the name \emph{dual} Josephson effect. We rule out its connection to the physics of Majorana boundary zero-modes and rather link it to the non-trivial topological properties of the two adiabatic pumps. 
Specifically, the effect survives to the shrinking of the central part to a symmetry defect, a point-like  region in space where the symmetries of the Thouless pump are violated. The non-trivial interplay between the topological bulk of the quantum pump and the defect are fully responsible for the dual Josephson effect. 

The dissipative pump of single electrons which appears in the presence of boundaries is turned into a dissipationless parity pump.
This observation opens the path to generalizations to higher-order magneto-Josephson effects, e.g.\ with periodicity of $8 \pi$. We discussed this phenomenology for several sizes of the junction and using different techniques.
Besides reinterpreting the magneto-Josephson effect originally developed in the context of superconducting nanowires and Majorana modes, our discussion shows the importance of the interplay between the bulk and edge properties of a topological quantum system in determining the final behaviour of the device. 
In particular, this is a realistic example where breaking the symmetries that underlie the topological description of the adiabatic quantum pump does not completely trivialize the system but leads to a novel and interesting phenomenology.

\appendix

\section{Models of Thouless pumps}
\label{App:Thouless:Pumps}

\subsection{Mapping onto a 2D system}\label{app:Harper_Hofstadter}

\begin{figure}
    \centering
    \includegraphics[width = \linewidth]{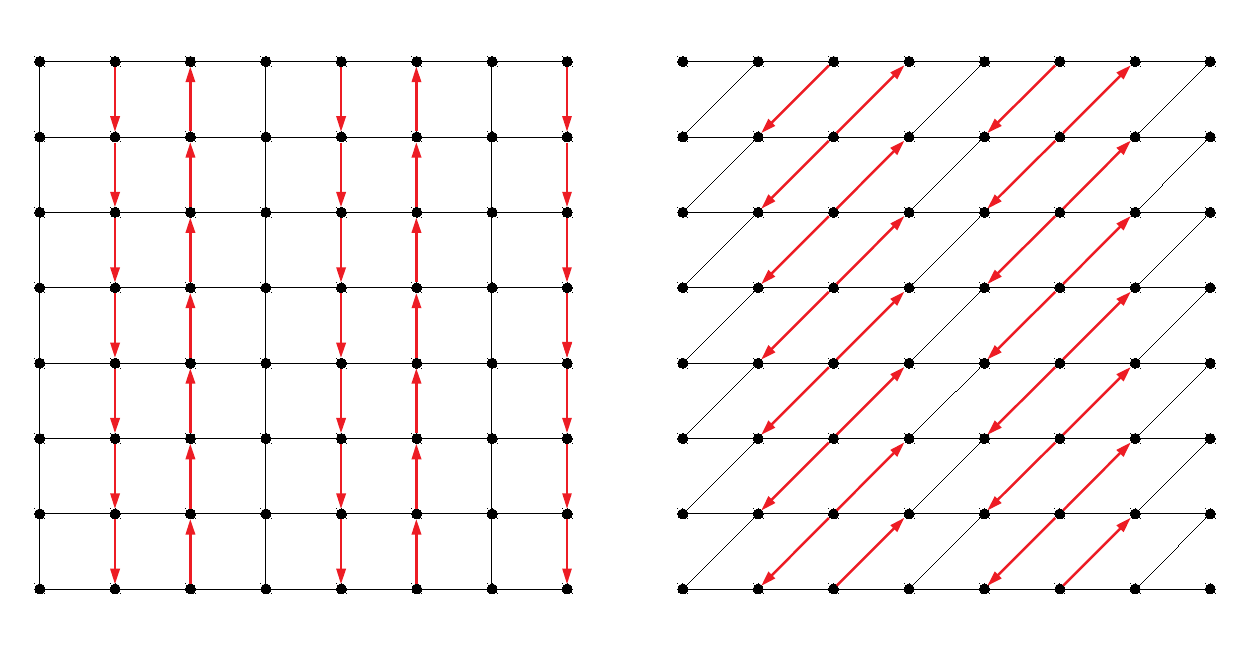}
    \caption{Sketch of Hamiltonians~\eqref{eq:H_pump_micro_2D} (left panel) and~\eqref{eq:H_pump_micro_phase_2D} (right panel), where black lines indicate real hopping terms and red arrows carry a phase $-\frac{2 \pi}{3}$. They respectively form a 2D square lattice and a 2D oblique lattice with magnetic flux $\frac{2\pi}{3}$ per plaquette, that is a 2D Harper-Hofstadter model with flux $1/3$.}
    \label{fig:phases_1}
\end{figure}

1D Thouless pumps can generically be mapped onto 2D systems by considering the adiabatic parameter as a momentum along an artificial dimension $y$. Here, we apply this procedure to Hamiltonians~\eqref{eq:H_pump_micro}, \eqref{eq:H_pump_micro_phase} and~\eqref{eq:H_pump_micro_ampl} introduced in Sec.~\ref{sect:models}, with $\theta$ the adiabatic parameter; their 2D equivalents are:
\begin{subequations}
\begin{align}
    H_{\text{Harp}}^{2D} &= \sum_{x, \theta} \bigg\{ \left( -t \hat{a}^\dagger_{x, \theta} \hat{a}_{x+1, \theta} + \text{h.c.} \right) 
    \\
    \nonumber
    & \qquad \qquad - V_0 \cos\left(\frac{2\pi j}{3} + \theta\right) \bigg\}\, \hat{a}^\dagger_{x, \theta} \hat{a}_{x, \theta}
    \\
    H_1^{2D} &=  \sum_{x, \theta} - \left( t + t' e^{i\left(\frac{2 \pi j}{3} + \theta\right)}\right) \hat{a}^\dagger_{x, \theta} \hat{a}_{x+1, \theta} + \text{h.c.} 
    \\
    H_2^{2D} &=  \sum_{x, \theta} - \left( t + t' \cos\left(\frac{2 \pi j}{3} + \theta\right)\right)  \hat{a}^\dagger_{x, \theta} \hat{a}_{x+1, \theta} + \text{h.c.}
\end{align}
\end{subequations}
In real space, this corresponds to:
\begin{subequations}
\begin{align}
    \label{eq:H_pump_micro_2D}
    H_{\text{Harp}}^{2D} &= \sum_{x, y}  -t \hat{a}^\dagger_{x,y} \hat{a}_{x+1,y} - \frac{V_0}{2} e^{i\frac{2 \pi x}{3}} \hat{a}^\dagger_{x,y} \hat{a}_{x,y+1} + \text{h.c.} 
    \\
    \label{eq:H_pump_micro_phase_2D}
    H_1^{2D} &=  \sum_{x, y}  -t \hat{a}^\dagger_{x,y} \hat{a}_{x+1,y} - t' e^{i\frac{2 \pi x}{3}}\, \hat{a}^\dagger_{x,y} \hat{a}_{x+1,y+1} + \text{h.c.}
    \\
    \nonumber
    H_2^{2D} &=  \sum_{x, y}  -t \hat{a}^\dagger_{x,y} \hat{a}_{x+1,y} - \frac{t'}{2} e^{i\frac{2 \pi x}{3}}\, \hat{a}^\dagger_{x,y} \hat{a}_{x+1,y+1} 
    \\
    \label{eq:H_pump_micro_ampl_2D}
    & \qquad \qquad - \frac{t'}{2} e^{i\frac{2 \pi x}{3}}\, \hat{a}^\dagger_{x,y} \hat{a}_{x+1,y-1} + \text{h.c.}
\end{align}
\end{subequations}

Hamiltonians~\eqref{eq:H_pump_micro_2D} and~\eqref{eq:H_pump_micro_phase_2D} are sketched in Fig.~\ref{fig:phases_1}. They respectively describe a 2D square lattice and a 2D oblique lattice with magnetic flux $\frac{2\pi}{3}$ per plaquette, that is a 2D Harper-Hofstadter model with flux $1/3$. Their lowest band therefore have a Chern number $C=-1$.

\begin{figure}
    \centering
    \includegraphics[width = \linewidth]{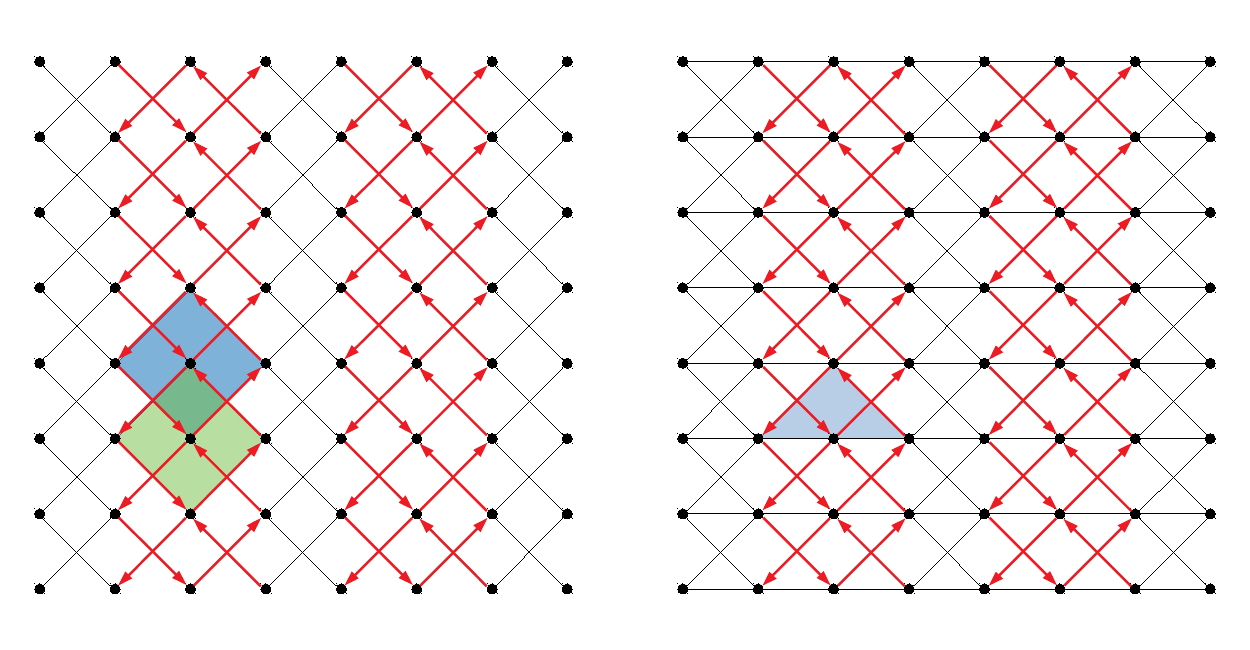}
    \caption{Sketch of Hamiltonian~\eqref{eq:H_pump_micro_ampl_2D}, where black lines indicate real hopping terms and red arrows carry a phase $-\frac{2 \pi}{3}$. 
    \textbf{Left panel:} For $t=0$, the system is made of two independent superimposed Harper Hofstadter models with flux $2/3$; the corresponding unit cells are shown in blue and green. \textbf{Right panel:} For large values of $t$, the two replicas are coupled, and their unit cells are split; the system turns into a Harper Hofstadter model with flux $1/3$ and triangular unit cells (shown in blue). The topological phase transition between these two regimes occurs at $t' = 4t$.}
    \label{fig:phases_2}
\end{figure}

Hamiltonian~\eqref{eq:H_pump_micro_ampl_2D} is sketched in Fig.~\ref{fig:phases_2}. For $t=0$, it corresponds to two independent superimposed Harper Hofstadter models with flux $2/3$, and its lowest band has a Chern number $C=2$. As $t$ is increased, both Harper Hofstadter models get coupled, and their unit cells are split in halves. A topological phase transition occurs at $t' = 4t$, and for $t'<4t$ the system can be understood as a Harper Hofstadter models with flux $1/3$ where the unit cells are triangular (see Fig.~\ref{fig:phases_2}); the Chern number of the lowest band is then $C=-1$. This topological phase transition is further discussed in Appendix~\ref{app:generalized_models}.

\subsection{Generalized models}\label{app:generalized_models}

\begin{figure}
    \centering
    \includegraphics[width = \linewidth]{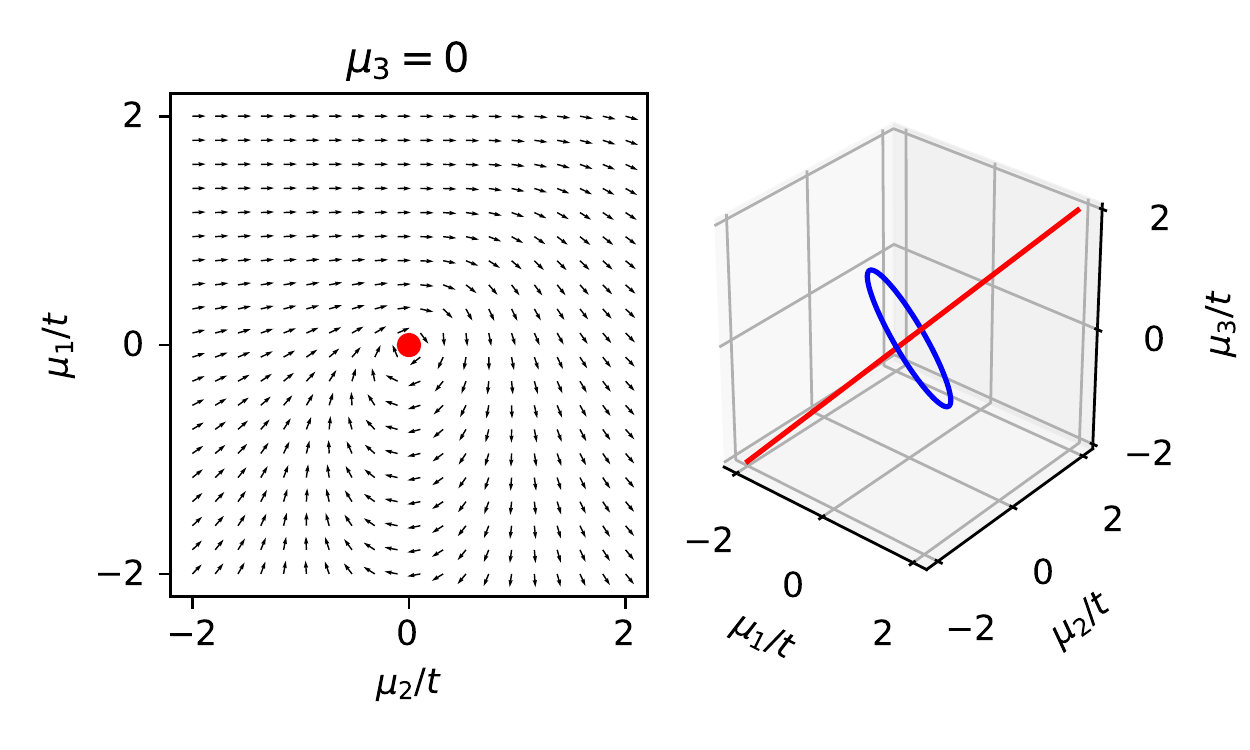}
    \caption{\textbf{Left panel:} Zak phase of Hamiltonian~\eqref{eq:H_pump_micro_gen} for $\mu_3 = 0$. The Zak phase has a vorticity $-1$ around the singular point $\mu_1 = \mu_2 = \mu_3 = 0$.
    \textbf{Right panel:} Singular line $\mu_1 = \mu_2 = \mu_3$ of Hamiltonian~\eqref{eq:H_pump_micro_gen} (shown in red). The blue circle corresponds to the closed loop obtained in parameter space the phase $\theta$ is winded from $-\pi$ to $\pi$ in the original model~\eqref{eq:H_pump_micro}; as it winds once around the singular line, its Chern number is $-1$. It was obtained for $V_0 = 1$, as in Fig.~\ref{fig:Pump_spectrum}.}
    \label{fig:3D_BI3_20}
\end{figure}

Hamiltonian~\eqref{eq:H_pump_micro} can be generalized by considering an arbitrary chemical potential with periodicity of $3$ sites:
\begin{equation}\label{eq:H_pump_micro_gen}
    H_{\text{Harp}}' = \sum_j \bigg\{ \left( -t \hat{a}^\dagger_j \hat{a}_{j+1} + \text{h.c.} \right) - \mu_j \hat{a}^\dagger_j \hat{a}_{j} \bigg\},
\end{equation}
where $\mu_j$ is 3-sites periodic. The model is described by $3$ real parameters $\mu_{1,2,3}$, and the gap between the two lowest bands only closes for $\mu_1 = \mu_2 = \mu_3$, which is a line in parameter space. Along this singular line, the Zak phase has a nonzero vorticity, as shown in Fig.~\ref{fig:3D_BI3_20}. For an adiabatic closed loop in parameter space, the corresponding Chern number can be computed as the winding number of the Zak phase; here, this corresponds to the number of times this closed loop winds around the singular line. In our original model~\eqref{eq:H_pump_micro}, winding the phase $\theta$ from $-\pi$ to $\pi$  draws circle around the singular line, as shown in Fig.~\ref{fig:3D_BI3_20}; the corresponding Chern number is therefore $C = -1$. 

\begin{figure}
    \centering
    \includegraphics[width = \linewidth]{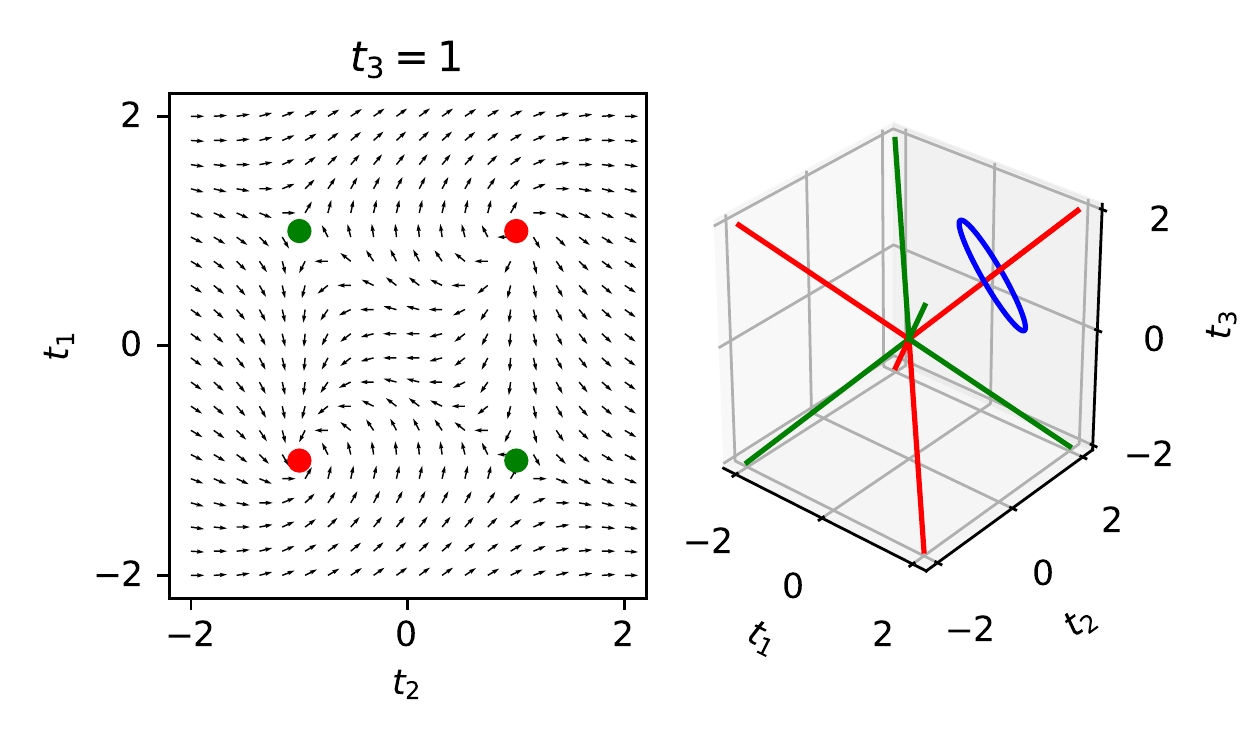}
    \caption{\textbf{Left panel:} Zak phase of Hamiltonian~\eqref{eq:H_pump_ampl_gen} for $t_3 = 1$. The Zak phase has a vorticity $+1$ (green) or $-1$ (red) around the singular points at $|t_1| = |t_2| = |t_3| = 1$.
    \textbf{Right panel:} Singular lines $|t_1| = |t_2| = |t_3|$ of Hamiltonian~\eqref{eq:H_pump_ampl_gen} with different vorticities for the Zak phase (shown in green and red). The blue circle corresponds to the closed loop obtained in parameter space the phase $\theta$ is winded from $-\pi$ to $\pi$ in Hamiltonian~\eqref{eq:H_pump_micro_ampl}. If $t'<4t$, it winds once around singular line with vorticity $-1$, and the corresponding Chern number is $-1$; if $t'>4t$, it also encircles three singular line with vorticity $+1$, and the corresponding Chern number is $2$. It was obtained for $t' = 0.8t$, as in Fig.~\ref{fig:Pump_spectrum_2}.}
    \label{fig:3D_Ampl3}
\end{figure}

With the same idea, we can study a system arbitrary hopping terms with periodicity of $3$ sites:
\begin{equation}\label{eq:H_pump_ampl_gen}
    H' = \sum_j \bigg\{-t_j \hat{a}^\dagger_j \hat{a}_{j+1} + \text{h.c.} \bigg\},
\end{equation}
where $t_j$ is 3-sites periodic. The model is described by $3$ complex parameters $t_{1,2,3}$, and the gap between the two lowest bands only closes for $|t_1| = |t_2| = |t_3|$. If the hopping amplitudes are real, the singular points are the diagonals of the unit cube in the parameter space, and they carry different vorticities ($\pm 1$) for the Zak phase, as shown in Fig.~\ref{fig:3D_Ampl3}. Winding the phase $\theta$ from $-\pi$ to $\pi$ in Hamiltonian~\eqref{eq:H_pump_micro_ampl} draws circle around one (if $t'<4t$) or four (if $t'>4t$) of these singular lines, which amounts to a Chern number $C = -1$ or $C = 2$ for the lowest band (See Fig.~\ref{fig:3D_Ampl3}).

\section{Transfer matrix approach}\label{app:transfer_matrix}

Let us consider the Pump--Kitaev--Pump configuration; we develop the field operators on an eigenbasis of the Hamiltonian as:
\begin{equation}
	\left(
	\begin{array}{c}
		\psi_R(x) \\
		\psi_L(x) \\
		\psi_L^\dagger(x) \\
		-\psi_R^\dagger(x) 
	\end{array}
	\right)
	= \sum_n 
	\left(
	\begin{array}{c}
		A_n(x) \\
		B_n(x) \\
		C_n(x) \\
		D_n(x) 
	\end{array}
	\right)
	\hat \eta_n + 
	\left(
	\begin{array}{c}
		-D_n^*(x) \\
		C_n^*(x) \\
		B_n^*(x) \\
		-A_n^*(x)
	\end{array}
	\right)
	\hat \eta_n^\dagger,
\end{equation}
and the equation for the wavefunctions in the 1D Harper model and Kitaev region read:
\begin{equation}
    h_{\rm 1D Harper, K} 
    \left(
	\begin{array}{c}
		A_n(x) \\
		B_n(x) \\
		C_n(x) \\
		D_n(x) 
	\end{array}
	\right)
	= \varepsilon_n
	\left(
	\begin{array}{c}
		A_n(x) \\
		B_n(x) \\
		C_n(x) \\
		D_n(x) 
	\end{array}
	\right).
\end{equation}
They can be solved independently in each region, and the energy of a mode is fixed by a continuity equation at the boundaries. 

For $\varepsilon \ll V_0, \Delta$, that is, for $L\gg 1$, these continuity equations break down to a reflection coefficient at the boundaries of the Kitaev region, and the two pumps can be disregarded. More precisely, a solution in the Kitaev region reads:
\begin{align}
    A(x) & = a_1 e^{\frac{\Delta}{v_F} x} + a_2 e^{-\frac{\Delta}{v_F} x}\\
    \nonumber
    C(x) & = a_1 e^{-i \beta_\Delta} e^{\frac{\Delta}{v_F} x} + a_2 e^{i \beta_\Delta} e^{-\frac{\Delta}{v_F} x}\\
    \nonumber 
    D(x) & = a_3 e^{\frac{\Delta}{v_F} x} + a_4 e^{-\frac{\Delta}{v_F} x}\\
    \nonumber
    B(x) & = a_3 e^{-i \beta_\Delta} e^{\frac{\Delta}{v_F} x} + a_4 e^{i \beta_\Delta} e^{-\frac{\Delta}{v_F} x},
\end{align}
where $\beta_\Delta = \frac{\pi}{2}+\frac{\varepsilon}{\Delta}$, $a_{1\hdots4}$ are integration constants, and the superconducting phase $\phi$ has been set to zero. One can compute the transfer matrix of the system, which has a simple form since components $A, C$ and $D,B$ are decoupled. One finds:
\begin{equation}
	\left(
	\begin{array}{c}
		A(L) \\
		C(L) 
	\end{array}
	\right)
	= M
	\left(
	\begin{array}{c}
		A(0) \\
		C(0) 
	\end{array}
	\right),
\end{equation}
(idem for $D,B$), with:
\begin{equation}
    M = \frac{1}{\sin( \beta_\Delta)}
    \left(
    \begin{array}{cc}
         \sin\left( \beta_\Delta - i \frac{\Delta}{v_F}L \right) & i \sinh \left(\frac{\Delta}{v_F}L\right) \\
         -i \sinh \left( \frac{\Delta}{v_F}L \right)& \sin\left( \beta_\Delta + i \frac{\Delta}{v_F}L \right) \\
    \end{array}
    \right)
\end{equation}

Let $\theta_{\rm left, right}$ be the pump parameters in the left and right region; then, the boundary conditions in $x=0,L$ are:
\begin{align}
	\left(
	\begin{array}{c}
		A(0) \\
		C(0) 
	\end{array}
	\right)
	& = e^{i \theta_{\rm left}} R_{\beta_V}
	\left(
	\begin{array}{c}
		D(0) \\
		B(0) 
	\end{array}
	\right)
	\\
	\nonumber
	\left(
	\begin{array}{c}
		A(L) \\
		C(L) 
	\end{array}
	\right)
	& = - e^{i \theta_{\rm right}} R_{-\beta_V}
	\left(
	\begin{array}{c}
		D(L) \\
		B(L) 
	\end{array}
	\right),
\end{align}
where 
\begin{equation}
    R_\alpha = 
    \left(
    \begin{array}{cc}
		0 & e^{i\alpha}\\
		e^{-i\alpha} & 0 
	\end{array}
	\right),
	\qquad
	R_\alpha^2 = I,
	\qquad
	\beta_V = \frac{\pi}{2} + \frac{\varepsilon}{V}
\end{equation}

They lead to the consistency equation:
\begin{equation}
	\left(
	\begin{array}{c}
		A(0) \\
		C(0) 
	\end{array}
	\right)
	=
	e^{i(\theta_{\rm left} - \theta_{\rm right})} R_{\beta_V} M^{-1} R_{-\beta_V} M
	\left(
	\begin{array}{cc}
		A(0) \\
		D(0) 
	\end{array}
	\right),
\end{equation}
which can be interpreted as a Bohr-Sommerfeld condition on the energy. It leads to:
\begin{equation}
    \text{det}(I-Q) = 0, \qquad Q = e^{i(\theta_{\rm left} - \theta_{\rm right})} R_{\beta_V} M^{-1} R_{-\beta_V} M
\end{equation}
In the limit $L\gg1$, it is solved by:
\begin{equation}\label{spectrum}
    \varepsilon = \pm \frac{2 \Delta V}{\Delta + V} e^{-\frac{\Delta}{v_F} L} \cos \left(\frac{\theta_{\rm left} - \theta_{\rm right}}{2}\right),
\end{equation}
where the characteristic length of the exponential decay is the coherence length $\xi = \hbar v_F/\Delta$ of the superconductor.

\section{Bosonization approach to the fractional Josephson effect} \label{App:8PI:Bos}

We provide a derivation to the fractional Josephson effects within the framework of bosonization. This gives an alternative computation of the $4 \pi$ Josephson effect which can be extended to higher periodicity as bosonization naturally accounts for non-quadratic terms in the Hamiltonian.

\subsection{Harper-Kitaev-Harper system}
Let us consider the geometry of the dual Josephson effect, introduced in Sec.~\ref{sect:duality}. For the central Kitaev chain, see Eq.~\eqref{eq:H_kitaev_continuum}, the bosonized Hamiltonian is
\begin{equation}\label{bos}
H_{bos} =   \int_0^L dx \left[ \frac{v_F}{2 \pi} \left[ (\partial_x \hat \theta)^2 + (\partial_x \hat \varphi)^2 \right]  - \frac{\Delta}{a_0}  \cos (2 \hat \theta ) \right],
\end{equation}
with the lattice spacing $a_0$. The conjugate variables $\hat \varphi$ and $\hat \theta$, satisfying $[ \hat \varphi (x) ,  \partial_x \hat \theta(x') ] = i \pi \delta (x-x')$, describe the long-wavelength excitations of the left and right movers via 
\begin{equation}
 \psi_{R/L} (x) = \frac{1}{\sqrt{2 \pi a_0}} e^{i [\hat \theta(x) \pm \hat \varphi (x)]}.
\end{equation}
The outer regions, $x<0$ and $x >L$, are respectively described by two 1D Harper models~\eqref{eq:H_pump_continuum}, or Thouless pumps, with different phases $\theta_{\rm left}$ and $\theta_{\rm right}$. Without loss of generality, we assume very large gaps in the Thouless pumps which impose the low-energy boundary conditions (BC)
\begin{equation}
 \psi_{R} (0) = i e^{i \theta_{\rm left}} \psi_{L} (0), \quad \psi_{R} (L) = -i e^{i \theta_{\rm right}} \psi_{L} (L).
\end{equation}
The BC can be view as the dual of Andreev reflections taking place at the two ends. They are consistent with the mode expansion
\begin{equation}\label{mode-exp}
\begin{split}
\hat \varphi (x)  & =  \varphi_0 + \frac{x}{L} \left( \pi \hat N_0 - \delta \right) +  \hat \varphi_e (x) \\[2mm]
\hat \theta (x)  & = \hat \theta_0 +  \hat \theta_e (x), 
\end{split}
\end{equation}
with $ \varphi_0 = \pi/4 + \theta_{\rm left}/2$ and $2 \delta = \theta_{\rm left} - \theta_{\rm right}$. We have singled out the conjugate zero-mode operators satisfying $[ \hat N_0, \hat \theta_0] = i$. $ \hat \theta_0$ is an angle variable so that $ \hat \theta_0 \sim \hat \theta_0+2 \pi$ are identified. The rest of the expansion is 
\begin{equation}
\begin{split}
\hat \varphi_e (x) &=  \sum_{k > 0}   \sqrt{\frac{\pi}{L k}}  i  \sin (k x) ( \hat a_k^\dagger - \hat a_{k}), \\[1mm]
 \hat \theta_e (x)  &=  \sum_{k > 0}   \sqrt{\frac{\pi}{L k}}   \cos (k x) ( \hat a_k^\dagger + \hat a_{k})
\end{split}
\end{equation}
 with the ladder operators $ \hat a_k$ satisfying $[\hat a_k,\hat a_{k'}^\dagger] = \delta_{k,k'}$. The variables $ \hat a_k$, $ \hat a_k^\dagger$ are in fact gapped by the cosine potential~\cite{Lindner_Berg_2012} and, once integrated, they replace~\cite{Fidkowski2011} the lattice spacing $a_0$  by the superconductor coherence length $\xi$ in Eq.~\eqref{bos}.

\subsection{Cooper pair box analogy}

Keeping only the zero modes inserted in Eq.~\eqref{bos}, we find the effective low-energy Hamiltonian
\begin{equation}\label{zeromodeha}
  H_{\delta} = \frac{v_F}{2 \pi L} \left(\pi \hat N_0 - \delta \right)^2 - \frac{\Delta L}{\xi}  \cos ( 2 \hat \theta_0 )
\end{equation}
In the context of circuit quantum electrodynamics, $H_{\delta}$ is the Hamiltonian of a transmon and it can be exactly solved in terms of Mathieu functions~\cite{koch2007}. We switch to a first quantized form,  
\begin{equation}\label{zeromodeha2}
H_{\delta} = E_C (i \partial_\theta - \delta/\pi)^2 - E_J \cos  2 \theta ,
\end{equation}
with the charging energy $E_C = \pi v_F/2 L$, the Josephson energy $E_J = \Delta L/\xi$ and the eigenfunctions $\varphi (\theta)$, solving
\begin{equation}\label{eigenfunction}
  H_{\delta} \, \varphi (\theta)  = \varepsilon \,  \varphi (\theta), 
\end{equation}
must satisfy the periodicity condition $\varphi (\theta + 2 \pi) = \varphi (\theta)$. 

Let us consider for the moment the periodic Hamiltonian $H_{\delta=0}$ and drop the restriction on the periodicity of wavefunctions. Using Bloch's theorem, we note $\tilde \varphi_k (\theta)$ its eigenstate with quasimomentum $k \in [0,2 ]$. In principle,  $\tilde \varphi_k (\theta)$ also carries a band index but we omit it since we keep only the lowest-energy subband. The eigenstate can be written as  $\tilde \varphi_k (\theta) = e^{i k \theta} u_k (\theta)$. The  function $u_k$ is $\pi$-periodic and solves $H_{\delta=\pi k} u_k = \varepsilon_k u_k$ with eigenenergy $\varepsilon_k$. 

Coming back to our model with the eigenvalue problem~\eqref{eigenfunction}, we find that there are two low-energy solutions, $\varphi_j (\theta) = e^{-i  \delta \theta/\pi} \tilde \varphi_{\delta/\pi+j} (\theta)$, with $j=0,1$ and energies $\varepsilon_{\delta/\pi+j}$,  satisfying the required $2 \pi$-periodicity. For a long Kitaev chain $E_J \gg E_C$, or $L \gg \xi$, the energy dispersion of the lowest subband of the unconfined Hamiltonian $H_{\delta=0}$ takes the form $\varepsilon_k = E_0 + \Gamma \cos(\pi k)$ such that
\begin{equation}\label{energies}
\left. \varepsilon_{\delta/\pi+j} \right|_{j=0,1} =  \pm  \Gamma \cos \delta,
\end{equation}
where energies are measured with respect to $E_0 \sim -E_J$, similar to a tight-binding model with bandwidth $\Gamma \sim e^{-2 \sqrt{2 E_J/E_C}}$, $\sqrt{E_J/E_C} \sim L/\xi$. These energies then coincide with the spectrum~\eqref{spectrum} and exhibit a $4 \pi$ periodicity with respect to the phase difference $\theta_{\rm left} - \theta_{\rm right}$ as already anticipated.

Introducing the  Wannier function ${\mathcal W}_0 (\theta)$ in the lowest-energy band (${\mathcal W}_0$ becomes a gaussian centered at $\theta = 0$ for $E_J \gg E_C$), the two eigenstates take the suggestive form
\begin{equation}
\varphi_j (\theta)  = \sum_{p \in \mathbb{Z}} {\mathcal W}_0 (\theta - p \pi) e^{i \frac{\delta}{\pi} (p \pi - \theta)} e^{i p \pi j}
\end{equation}
describing the coupling between the different minima of the $\cos 2 \theta$ potential by instantons~\cite{Zhang_Kane_2014_PRL}. The parity operator $\hat P = (-1)^{\hat N_0}$ generates the shift $\theta \to \theta+\pi$, and
\begin{equation}
\hat P \varphi_j (\theta) = \varphi_j (\theta+\pi) = (-1)^{j} \varphi_j (\theta).
\end{equation}
Therefore the parity $\hat P$ is the symmetry protecting the energy crossing at $\delta = \pi/2$ in Eq.~\eqref{energies}. However, it is not the parity of the Kitaev chain given by $\hat P_K = \hat P e^{-i \delta}$. If we take a given state, {\it e.g.} $\varphi_0 (\theta)$, and advance $\delta$ adiabatically by $\pi$, then $\hat P$ is conserved but $\hat P_K \to -\hat P_K$ in agreement with the picture of the charge unloaded by the Thouless pump discussed in Sec.~\ref{sect:dual_effect_lattice_model}.

We note that the conventional $4 \pi$ Josephson effect arising in the Kitaev-Harper-Kitaev geometry is straightforwardly derived by simply exchanging the roles of the conjugate variables $\hat \theta$ and $\hat \varphi$ in the above analysis, corresponding to the dual transformation of Eq.~\eqref{eq:duality_mapping}.

\subsection{\texorpdfstring{$8\pi$ Josephson effect}{8 pi Josephson effect}}

It is straightforward to generalize the above discussion to a $\mathbb{Z}_4$-pairing central region, {\it i.e.} the Hamiltonian~\eqref{eq:H_quadruplet}, instead of the Kitaev chain. The initial bosonized Hamiltonian has the form of Eq.~\eqref{bos} with $\cos 4 \hat \theta$ replacing $\cos 2 \hat  \theta$. The reduction to the zero-energy modes then simply replaces $\cos 2 \theta$ by $\cos 4 \theta$ in Eq.~\eqref{zeromodeha2}. Therefore the only difference with the Kitaev chain is that the periodicity of the cosine potential is now $\pi/2$ instead of $\pi$. The functions $u_k (\theta)$ have also a reduced $\theta-$periodicity of $\pi/2$ and the size of the Brillouin zone is $4$ instead of $2$.
As a result, we obtain a low-energy subspace with four states
\begin{equation}
  \varphi_j (\theta) = e^{-i  \delta \theta/\pi} \tilde \varphi_{\delta/\pi+j} (\theta), \qquad j =0,1,2,3
\end{equation}
corresponding to the four distincts energies $\varepsilon_{\delta/\pi+j}$. In the large length limit, $L \gg \xi$, one finds an explicit form  $\varepsilon_k = E_0 + \Gamma \cos(\pi k/2)$ so that
\begin{equation}
  \left. \varepsilon_{\delta/\pi+j} \right|_{j=0,2} = \pm  \Gamma \cos \delta, \quad \left. \varepsilon_{\delta/\pi+j} \right|_{j=1,3} = \mp  \Gamma \sin \delta,
\end{equation}
when substracting $E_0$. The four states exhibit a cyclic spectral flow in  $\theta_{\rm left} - \theta_{\rm right}$
with an $8 \pi$ periodicity~\cite{Mazza_Mora_2018}. The energy crossing are protected by the generalized parity $\hat P = e^{i \pi \hat N_0/2}$.

Moreover, the $8 \pi$ Josephson effect mediated by the quantum spin Hall edge with time-reversal symmetry is directly obtained from the duality transformation $\hat \theta \leftrightarrow \hat \varphi$. We thus reproduce the results of Ref.~\onlinecite{Zhang_Kane_2014_PRL,kane2015,pedder2017}.

\bibliography{Josephson}

\end{document}